\documentclass[reprint,aps,pra,groupedaddress]{revtex4-2}

\usepackage{physics}
\usepackage{amsmath}
\usepackage{xcolor}
\usepackage{graphicx}
\usepackage[hidelinks]{hyperref}
\usepackage{cleveref}
\usepackage{float}
\usepackage[normalem]{ulem} 
\crefname{equation}{Eq.}{} 

\newcommand{\pati}[2]{}

\begin{document}

\title{Scattering and Chirping at Accelerated Interfaces}

\author{Klaas De Kinder}
\author{Amir Bahrami}
\author{Christophe Caloz}
\email[]{christophe.caloz@kuleuven.be}
\affiliation{Department of Electrical Engineering, KU Leuven, Leuven, 3000,  Belgium}

\date{\today}

\begin{abstract}
    Space-time varying media with moving interfaces unlock new ways to manipulate electromagnetic waves. Yet, analytical solutions have been mostly limited to interfaces moving at constant velocity or constant proper acceleration. Here, we present exact scattering solutions for an arbitrarily accelerating interface, derived directly in the laboratory frame through a suitable change of variables. We show that acceleration introduces rich effects that do not occur with uniform motion, including transitions between multiple velocity regimes, multiple scattering events and generalized frequency chirping. We also solve the inverse problem of designing an interface trajectory that produces a desired chirping profile, demonstrating how tailored acceleration can synthesize complex frequency modulations. These results provide a fundamental framework to understand and control wave interactions with accelerated boundaries, opening pathways for advanced applications in space-time signal processing and dynamic pulse shaping.
\end{abstract}

\maketitle

\section{Introduction}
    \pati{Definition GSTEMs}{
    }

    Space-time varying media are engineered structures whose electromagnetic properties, such as the refractive index, are modulated in both space and time~\cite{Caloz2019a_ST_Metamaterials_PUB,Caloz2019b_ST_Metamaterials_PUB}. A characteristic parameter is their modulation velocity which refers to the speed at which these parameters change in space and time. Depending on whether this velocity is infinite, constant or time-dependent, space-time varying media exhibit distinct wave interactions and enable new physical effects~\cite{Caloz2022_GSTEMs_PUB}. 

    \pati{TEMs}{
    }
    
    Purely temporal interfaces have been studied extensively ~\cite{Morgenthaler1958_TEM_PUB,Pendry2022_Review_TEM_PUB,Mostafa2024_PUB}. These unlock a broad range of phenomena unattainable by spatially static media, including, time reversal~\cite{Fink2016_TR_TEM_PUB}, operation beyond fundamental bounds~\cite{Monticone2024_Rozanov_Bound_TEM_PUB,Alu2019_Chu_Limit_TEM_PUB,Hadad2018_Bode-Fano_Bound_TEM_PUB}, photon cooling~\cite{Pendry2024_Air_Cond_Phot_USTEM_PUB}, inverse prism decomposition~\cite{Caloz2018_Inverse_Prism_PUB}, temporal impedance matching~\cite{Engheta2020_Aiming_PUB}, temporal aiming~\cite{Engheta2020_Coating_PUB}, parametric amplification~\cite{Tien1958_PUB}, beam splitting~\cite{Guerreiro2003_PUB}, photon generation~\cite{Guerreiro2000_PUB}, temporal Brewster refraction~\cite{Engheta2021_Brewster_PUB}, scattering at bianisotropic time interfaces~\cite{Tretyakov2024_PUB}, polarization conversion~\cite{Werner2021_PUB} and temporal analog Faraday rotation~\cite{Huanan2023_PUB,Alu2022_PUB}. 

    \pati{Moving Interfaces}{
    }
    
    Moving interfaces~\cite{Cassedy1963_PUB,Cassedy1967_PUB,Bolotovskii1972_PUB,Deck-Leger2019_Uni_Vel_PUB,Gaafar2019_PUB} have also attracted significant attention and support effects such as Doppler frequency shifting~\cite{Seftor1976_PUB,Lampe1978_PUB,Deck-Leger2019_Uni_Vel_PUB}, magnetless nonreciprocity~\cite{Sounas2024_PUB,Caloz2017_Nonreciprocity_PUB,Alu2014_PUB}, space-time reversal~\cite{Deck-Leger2018_PUB}, dynamic diffraction~\cite{Eleftheriades2019_PUB}, asymmetric bandgap isolation~\cite{Deck-Leger2019_Uni_Vel_PUB,Cassedy1963_PUB,Cassedy1967_PUB,Caloz2017_Isolation_PUB}, light deflection~\cite{Pendry2019_PUB,Pendry2021_Homogenization_PUB}, shock-wave generation~\cite{Joannopoulos2003_PUB}, Hawking radiation~\cite{Leonhardt2008_PUB,Leonhardt2019_PUB}, space-time Fresnel prism scattering~\cite{Li2023_Fresnel_Prism_PUB}, light amplification~\cite{Pendry2021_Gain_PUB,Luo2023_PUB}, space-time wedge multiple scattering~\cite{Bahrami2025_Wedges_USTEM_PUB} and Doppler pulse amplification~\cite{DeKinder2025_DoPA_USTEM}. These effects extend the capabilities and applications of purely temporal interfaces.

    \pati{ASTEM Gap}{
    }
    
    The aforementioned studies on moving interfaces are characterized by a constant modulation velocity. Recently, \emph{accelerated} interfaces have emerged, enabling complex temporal dynamics, enhanced spectral control and advanced functionalities, such as space-time lensing~\cite{Ostrovskii1975_Lens_PUB}, arbitrary pulse shaping~\cite{Bahrami2025_Pulse_Shap}, superluminal photon emission~\cite{Sloan2022_PUB}, frequency chirping~\cite{Bahrami2023_FDTD_PUB} and gravity analogs~\cite{Bahrami2023_ASTEMs_PUB}. However, a comprehensive analytical framework for such interfaces is still missing, except for the particular case of constant proper acceleration~\cite{Bahrami2023_ASTEMs_PUB,Bahrami2023_FDTD_PUB}. In this case, the problem can be addressed using the frame hopping method~\cite{Bladel2012_BOOK} with Rindler transformations~\cite{Misner1973_GR_BOOK}, but otherwise, when the acceleration has an arbitrary form, the frame hopping approach is inapplicable because no global coordinate transformation exists. Consequently, the fundamental problem of wave scattering at an arbitrarily accelerated interface remains unsolved, leaving the associated physics mostly unexplored. This gap limits practical applications, which remain scarce and underdeveloped.

    \pati{Contribution}{
    }

    In this work, we present, to the best of our knowledge, the first exact analytical scattering solutions for an arbitrary accelerated interface and elucidate the underlying physics. Instead of relying on the frame hopping approach, we solve the problem directly in the laboratory frame through a suitable change of variables, thereby avoiding the limitations of coordinate transformations. We show that acceleration introduces a time-dependent Doppler frequency shift, resulting in frequency chirping. We further address the synthesis problem, demonstrating how prescribed chirping profiles can be realized through tailored acceleration. These results open new possibilities for advanced wave manipulation and has potential applications in analog signal processing~\cite{Caloz2013_PUB} and chirped pulse shaping~\cite{Strickland1985_CPA_PUB}.

    \pati{Organization}{
    }
    
    The paper is structured as follows. Section~\ref{sec:Physics_of_Accelerated_Interfaces} outlines the fundamental physics governing wave scattering at arbitrarily accelerated interfaces. Section~\ref{sec:Analysis} derives the scattering solutions across all three velocity regimes and illustrates the results with four representative examples. Section~\ref{sec:Synthesis} addresses the synthesis problem and provides two examples. Finally, Sec.~\ref{sec:Conclusions} concludes the paper and discusses potential experimental realizations.

\section{Physics of Accelerated Interfaces}\label{sec:Physics_of_Accelerated_Interfaces}
    \begin{figure*}
        \centering
        \includegraphics[width=1.0\linewidth]{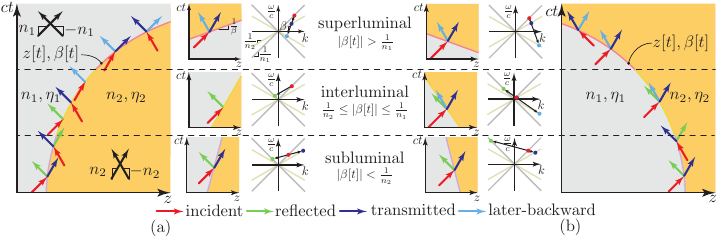} 
        \caption{Generalization of electromagnetic wave scattering from uniformly moving interfaces to accelerated interfaces, parametrized by a trajectory, $z{\left[t\right]}$, and normalized modulation velocity, $\beta{\left[t\right]} = \dd{z{\left[t\right]}}/\dd{(ct)}$. (a)~Left: space-time diagram of an accelerated interface, whose time-varying \emph{positive} velocity sweeps through all three velocity regimes---subluminal ($\left|\beta{\left[t\right]}\right| < 1/n_{2}$), interluminal ($1/n_{2} \leq \left|\beta{\left[t\right]}\right| \leq 1/n_{1}$) and superluminal ($\left|\beta{\left[t\right]}\right| > 1/n_{1}$), assuming $n_{1} < n_{2}$. Right: corresponding local constant-velocity reference cases for a forward-propagating wave in the first medium, each represented by the related space-time diagrams (left insets) and spectral transition diagrams (right insets). (b)~Same as (a) but for a \emph{negative} modulation velocity.}
        \label{fig:Overview}
    \end{figure*}

    \pati{Single Interface as Building Block}{
    }

    A single interface is the most fundamental building block of space-time varying media. This elementary structure, which may be called a space-time meta-atom~\cite{Horsley2023_Eigenpulses_PUB}, governs the basic interactions between electromagnetic waves and space-time modulations. More complex configurations, such as slabs, stacks, crystals and gradient structures, can be understood as a succession of such elementary interfaces. Even on their own, single interfaces enable practical functionalities such as space-time focusing~\cite{Ostrovskii1975_Lens_PUB} and pulse shaping~\cite{Bahrami2025_Pulse_Shap}.

    \pati{Problem Statement}{
    }

    We shall next describe the physics of wave scattering at accelerated interfaces with the help of Fig.~\ref{fig:Overview}. The interface separates two media with refractive indices $n_{1,2}$ and impedances $\eta_{1,2}$. Its trajectory is given by $z{\left[t\right]}$ with normalized modulation velocity $\beta{\left[t\right]} = v_{\text{m}}{\left[t\right]}/c = \dd{z{\left[t\right]}}/\dd{(ct)}$, both parametrized with laboratory time, $t$. The interface parametrization must be differentiable, which is typically satisfied for physically realizable interfaces; we can therefore consider that \emph{arbitrary physical} motion is allowed. The interface is illuminated by an incoming electromagnetic wave which, upon interaction, scatters and generates multiple components. The characteristics of these scattered waves depend on the instantaneous velocity of the interface at the time  of interaction. Our aim is to determine the scattered waveforms and their frequency content.

    \pati{Locally ASTEMs are USTEMs}{
    }

    Locally, that is, over a sufficiently short interaction region, an accelerated interface (curved trajectory) behaves like a uniformly moving interface (straight trajectory), a case that has been extensively studied~\cite{Deck-Leger2019_Uni_Vel_PUB}. This is analogous to the principle that, locally, general relativity reduces to special relativity~\cite{Misner1973_GR_BOOK}. This local equivalence offers an intuitive and practical approximation: well-established solutions for uniformly moving interfaces can be used to approximate the scattering behavior of accelerated interfaces over a small region. This approximation holds when the curvature of the trajectory is negligible over the interaction region, meaning the interface velocity varies slowly compared to the wave dynamics. This equivalence is illustrated by comparing the curved accelerated trajectory with its locally tangent uniform-velocity counterpart in the middle configurations of Fig.~\ref{fig:Overview}.

    \pati{Overview Scattering Configurations (locally)}{
    }

    Figure~\ref{fig:Overview} summarizes the different possible local scattering scenarios. These corresponding configurations fall into one of three velocity regimes, each with distinct physical behavior and determined by the value of the interface velocity at the time of scattering: subluminal (bottom part in Fig.~\ref{fig:Overview}), superluminal (top) and interluminal (middle), with $n_{1} < n_{2}$, which is assumed throughout the text. In the \emph{subluminal} regime ($\left|\beta{\left[t\right]}\right| < 1/n_{2}$), the interface velocity is smaller than the speed of light in both media. This case resembles that of a stationary interface: an incident wave generates a reflected wave in the incident medium, propagating in the opposite direction, and a transmitted wave in the other medium, propagating in the same direction as the incident wave. In the \emph{superluminal} regime ($\left|\beta{\left[t\right]}\right| > 1/n_{1}$), the interface velocity exceeds the speed of light in both media and the behavior is analogous to that of a purely temporal interface. In contrast to the subluminal case, the `reflected’ wave now appears in the opposite medium and is therefore often referred to as the `later-backward' wave. Between these regimes lies the \emph{interluminal} regime ($1/n_{2} \leq \left|\beta{\left[t\right]}\right| \leq 1/n_{1}$), the least understood so far, where the interface velocity is between the speeds of light in the two media. In this regime, the scattering behavior depends not only on whether the interface moves codirectionally or contradirectionally relative to the incident wave but also on the incident medium~\cite{Ostrovskii1967_Inter_PUB,Deck-Leger2019_Inter_CONF}. For instance, a forward codirectional wave incident from the first medium produces only a reflected wave (Fig.~\ref{fig:Overview}a), because although such a wave is fast enough to catch up with the interface ($1/n_{1}>\beta{\left[t\right]}$), it is too slow to propagate in the second medium ($1/n_{2}<\beta{\left[t\right]}$). In the contradirectional case (Fig.~\ref{fig:Overview}b), the same incident wave generates three scattered waves, since the interface now moves slower than light in the second medium ($\beta{\left[t\right]} < -1/n_{2} < 1/n_{2} $), producing, in addition to the reflected wave, a transmitted ($1/n_{2}$)---or later-forward---wave and a later-backward ($-1/n_{2}$) wave. Finally, there is never interluminal scattering from the second medium into the first medium when the modulation velocity is negative (Fig.~\ref{fig:Overview}b), because any wave in the second medium---whether codirectional or contradirectional---cannot catch up with the interface.

    \pati{Consequences from Acceleration (globally)}{
    }

    While interfaces moving at constant velocity are well understood, accelerating interfaces introduce complexities that cannot be captured by the local constant-velocity approximation alone. First, waves with some temporal duration may interact with the accelerating interface at different times and, therefore, at different velocities. When the wave encounters the interface at a point where the velocity regime changes, different portions of the wave undergo distinct interactions depending on the local velocity regime, resulting in more complex and diverse scattering dynamics. In contrast, a uniformly moving interface remains confined to a single velocity regime. Second, acceleration can lead to multiple scattering events. As the interface speeds up or down, it may overtake previously scattered waves, involving additional scattering. For example, in Fig.~\ref{fig:Overview}a, an initially subluminal scattering can be followed by a second interaction when the accelerating interface catches up with the transmitted wave. Third, acceleration induces frequency chirping. Unlike a uniformly moving interface, which produces constant Doppler shifting, an accelerating interface generates time-varying frequency shifting. This is evident from the frequency transition diagrams in Fig.~\ref{fig:Overview}: for constant velocity, the transition slope is fixed, resulting in fixed Doppler shifting, but when the velocity changes due to acceleration, the slope varies over time. Consequently, different frequency transitions occur at different times, producing frequency chirping in the scattered wave.

\section{Analysis}\label{sec:Analysis}
    \subsection{Scattering Formulas}\label{subsec:Scattering_Formulas}
    
    \pati{Assumptions}{
    }

    We consider two isotropic, linear and nondispersive media separated by a moving interface in a one-dimensional configuration. The interface moves along the $z$-direction and the electromagnetic fields depend solely on $z$ and $t$. The electric field is polarized along the $x$-direction and the magnetic field along $y$. General forward and backward traveling waveforms in each medium are expressed as 
    \begin{subequations}\label{eq:General_Waveforms}                
        \begin{align}
            E_{i}^{\pm} &= \psi_{i}^{\pm}{\left[\phi_{i}^{\pm}\right]}\,, 
            &H_{i}^{\pm} &= \pm \frac{1}{\eta_{i}}\psi_{i}^{\pm}{\left[\phi_{i}^{\pm}\right]}\,, \\
            D_{i}^{\pm} &= \frac{1}{u_{i}\eta_{i}}\psi_{i}^{\pm}{\left[\phi_{i}^{\pm}\right]}\,, 
            &B_{i}^{\pm} &= \pm \frac{1}{u_{i}}\psi_{i}^{\pm}{\left[\phi_{i}^{\pm}\right]}\,, 
        \end{align}
    \end{subequations}
    where square brackets, $\left[\cdot\right]$, indicate the arguments of the wave functions, $\psi_{i}^{\pm}$, and $\phi_{i}^{\pm} = z/u_{i} \mp t$ denote the traveling wave variables. The subscript, $i = 1,2$, labels the medium while the superscript, $\pm$, denotes forward ($+$) or backward ($-$) propagating waves. The wave speed in medium $i$ is given by $u_{i} = c/n_{i}$. In the following, we analyze the scattering of an electromagnetic pulse propagating from medium~$1$ to medium~$2$ across the interface, for all three velocity regimes (Fig.~\ref{fig:Overview}).

    \pati{Direct Space-Time Solution}{
    }
    
    Conventionally, moving interfaces are treated by transforming the problem to the comoving frame, where the boundary appears stationary, e.g., ~\cite{Lampe1978_PUB,Pendry2019_PUB,Pendry2021_Homogenization_PUB,Caloz2019b_ST_Metamaterials_PUB,Deck-Leger2019_Uni_Vel_PUB,Bahrami2023_ASTEMs_PUB}, an approach known as frame-hopping~\cite{Bladel2012_BOOK}. In this approach, Lorentz transformations are applied in the case of a constant velocity~\cite{Deck-Leger2019_Uni_Vel_PUB} and Rindler transformations in the case of a constant proper acceleration~\cite{Bahrami2023_ASTEMs_PUB}. Here, we solve the problem directly in the laboratory frame by enforcing \emph{moving} boundary conditions at the interface. The solutions are extended to the entire space-time domain, through a suitable change of variables, without requiring frame hopping.

    \subsubsection{Subluminal}

    \pati{Solution Boundary Conditions}{
    }

    For an incident forward-propagating pulse in the first medium, $\psi_{1}^{+}$, the reflected and transmitted waves are $\psi_{1}^{-}$ and $\psi_{2}^{+}$, respectively (Fig.~\ref{fig:Overview}). The boundary conditions require the continuity of $\boldsymbol{E} + c\boldsymbol{\beta} \times \boldsymbol{B}$ and $\boldsymbol{H} - c\boldsymbol{\beta} \times \boldsymbol{D}$ at the interface~\cite{Caloz2019b_ST_Metamaterials_PUB}. Feeding the general waveforms [Eqs.~\eqref{eq:General_Waveforms}] into these boundary conditions gives a system of equations for $\psi_{1}^{-}$ and $\psi_{2}^{+}$, whose solution reads (Sec.~\ref{subsec:appendix:Subluminal_Regime})
    \begin{subequations}\label{eq:Subluminal_Solution_System_of_Equations}
        \begin{align}
            \psi_{1}^{-}{\left[f_{1}^{-}{\left[t^{\star}\right]}\right]} &= \frac{\eta_{2}-\eta_{1}}{\eta_{2}+\eta_{1}}\frac{1-n_{1}\beta{\left[t^{\star}\right]}}{1+n_{1}\beta{\left[t^{\star}\right]}}\psi_{1}^{+}{\left[f_{1}^{+}{\left[t^{\star}\right]}\right]} \,, \\
            \psi_{2}^{+}{\left[f_{2}^{+}{\left[t^{\star}\right]}\right]} &= \frac{2\eta_{2}}{\eta_{2}+\eta_{1}}\frac{1-n_{1}\beta{\left[t^{\star}\right]}}{1-n_{2}\beta{\left[t^{\star}\right]}}\psi_{1}^{+}{\left[f_{1}^{+}{\left[t^{\star}\right]}\right]}\,,
        \end{align}
    \end{subequations}
    where $t^{\star}$ denotes the time at which the incident pulse scatters at the interface, referred to as the \emph{scattering time}. The auxiliary function $f_{i}^{\pm}$ is defined as
    \begin{equation}\label{eq:General_Definition_Function_f}
        f_{i}^{\pm}{\left[t^{\star}\right]} = \frac{z{\left[t^{\star}\right]}}{u_{i}} \mp t^{\star}\,.
    \end{equation}

    \pati{Extension to Space-Time Plane}{
    }

    Equations~\eqref{eq:Subluminal_Solution_System_of_Equations} define the scattered fields at the interface in terms of the scattering time, $t^{\star}$. To extend them across the entire space-time domain, the scattered waveforms must be re-expressed in terms of the traveling variables. This is accomplished by equating the arguments in Eqs.~\eqref{eq:Subluminal_Solution_System_of_Equations} to $\phi_{i}^{\pm}$, whose inverse is (Sec.~\ref{subsec:appendix:Subluminal_Regime}) 
    \begin{equation}\label{eq:Analysis_Transformation}
        t^{\star} \mapsto \left(f_{i}^{\pm}\right)^{-1}{\left[\phi_{i}^{\pm}\right]}\,.
    \end{equation}
    Equation~\eqref{eq:Analysis_Transformation} maps each space-time point $\left(z,ct\right)$---contained in the traveling wave variable $\phi_{i}^{\pm}$--- of the scattered wave in the entire space-time, back to its scattering time $t^{\star}$~\footnote{This scattering time may be interpreted as the emission time of the Huygens' source on the interface that produced the scattered field at $\left(z,ct\right)$.}. Inserting Eq.~\eqref{eq:Analysis_Transformation} into Eqs.~\eqref{eq:Subluminal_Solution_System_of_Equations} yields the subluminal field solutions
    \begin{subequations}\label{eq:Analysis_Subluminal_Solutions}
        \begin{align}
            &\psi_{1}^{-}{\left[\phi_{1}^{-}\right]} = \frac{\eta_{2}-\eta_{1}}{\eta_{2}+\eta_{1}}\frac{1-n_{1}\beta{\left[\left(f_{1}^{-}\right)^{-1}{\left[\phi_{1}^{-}\right]}\right]}}{1+n_{1}\beta{\left[\left(f_{1}^{-}\right)^{-1}{\left[\phi_{1}^{-}\right]}\right]}} \nonumber\\
            &\hspace{3.5cm} \psi_{1}^{+}{\left[f_{1}^{+}{\left[\left(f_{1}^{-}\right)^{-1}{\left[\phi_{1}^{-}\right]}\right]}\right]} \,, \label{eq:Analysis_Subluminal_Solutions_Reflection}\\
            &\psi_{2}^{+}{\left[\phi_{2}^{+}\right]} = \frac{2\eta_{2}}{\eta_{2}+\eta_{1}}\frac{1-n_{1}\beta{\left[\left(f_{2}^{+}\right)^{-1}{\left[\phi_{2}^{+}\right]}\right]}}{1-n_{2}\beta{\left[\left(f_{2}^{+}\right)^{-1}{\left[\phi_{2}^{+}\right]}\right]}} \nonumber \\
            &\hspace{3.5cm} \psi_{1}^{+}{\left[f_{1}^{+}{\left[\left(f_{2}^{+}\right)^{-1}{\left[\phi_{2}^{+}\right]}\right]}\right]}\,. \label{eq:Analysis_Subluminal_Solutions_Transmission}
        \end{align}
    \end{subequations}
    Each solution consists of three components. The first factor corresponds to the stationary Fresnel scattering coefficients ($\beta{\left[t\right]} = 0$). The second factor accounts for the Doppler-related stretching or compression factor. The last term is the incident wave retimed according to Eq.~\eqref{eq:Analysis_Transformation}.

    \pati{Chirping}{
    }

    The frequency shift, $\omega_{i}^{\pm}$, may be obtained by differentiating the phase of each scattered waveform with respect to $t$ (Sec.~\ref{subsec:appendix:Subluminal_Regime}), viz.,
    \begin{equation}\label{eq:Doppler_Shift}
        \omega_{i}^{\pm}{\left[t^{\star}\right]} = \pm\frac{n_{1}\beta{\left[t^{\star}\right]} - 1}{n_{i}\beta{\left[t^{\star}\right]} \mp 1}\,.
    \end{equation}
    For a non-constant interface velocity, $\beta{\left[t\right]}$, Eq.~\eqref{eq:Doppler_Shift} yields a time-varying Doppler shift, resulting in \emph{chirping} of the scattered wave. This expression recovers the well-known result for a constant-velocity interface~\cite{Deck-Leger2019_Uni_Vel_PUB} upon removing the $t^\star$ dependence, but generalizes it to arbitrary time-dependent motion through the scattering time dependency.

    \pati{Example USTEM}{
    }

    To illustrate Eqs.~\eqref{eq:Analysis_Subluminal_Solutions}, we consider an interface moving at constant normalized velocity, $\beta_{0}$,  parameterized as $z{\left[t\right]} = \beta_{0}ct$. In this case, the auxiliary function $f_{i}^{\pm}{\left[t^{\star}\right]} = \left(n_{i}\beta_{0}\mp 1\right)t^{\star}$ is linear due to the linearity of the equation of motion and can be easily inverted as $\left(f_{i}^{\pm}\right)^{-1}{\left[\phi_{i}^{\pm}\right]} = 1/\left(n_{i}\beta_{0}\mp 1\right)\phi_{i}^{\pm}$. Substituting this into Eqs.~\eqref{eq:Analysis_Subluminal_Solutions} yields
    \begin{subequations}
        \begin{align}
            &\psi_{1}^{-}{\left[\phi_{1}^{-}\right]} = \frac{\eta_{2}-\eta_{1}}{\eta_{2}+\eta_{1}}\frac{1-n_{1}\beta_{0}}{1+n_{1}\beta_{0}} \psi_{1}^{+}{\left[- \frac{1-n_{1}\beta_{0}}{1+n_{1}\beta_{0}}\phi_{1}^{-}\right]} \,, \\
            &\psi_{2}^{+}{\left[\phi_{2}^{+}\right]} = \frac{2\eta_{2}}{\eta_{2}+\eta_{1}}\frac{1-n_{1}\beta_{0}}{1-n_{2}\beta_{0}} \psi_{1}^{+}{\left[\frac{1-n_{1}\beta_{0}}{1-n_{2}\beta_{0}}\phi_{2}^{+}\right]}\,.
        \end{align}
    \end{subequations}
    As expected, both the reflected and the transmitted waves undergo a constant Doppler shift (factor multiplying the traveling waveform) because the interface velocity is constant, $\beta{\left[t^{\star}\right]} = \beta_{0}$, in agreement with Eq.~\eqref{eq:Doppler_Shift}. Consequently, the scattered waves do not exhibit chirping. Note that for more general trajectory profiles, Eq.~\eqref{eq:Analysis_Transformation} may not be analytically tractable and typically requires a numerical solution.

    \subsubsection{Superluminal}
    \pati{Methodology \& Solution}{
    }

    The approach to the superluminal regime follows the same procedure as in the subluminal case. For a forward-propagating incident wave in the first medium, the scattered waves are the later-backward wave, $\psi_{2}^{-}$ and transmitted wave, $\psi_{2}^{+}$. Once again, the boundary conditions enforce the continuity of $\boldsymbol{E} + c\boldsymbol{\beta} \times \boldsymbol{B}$ and $\boldsymbol{H} - c\boldsymbol{\beta} \times \boldsymbol{D}$ at the interface~\cite{Caloz2019b_ST_Metamaterials_PUB}. Inserting the general waveforms [Eqs.~\eqref{eq:General_Waveforms}] into the boundary conditions, solving the resulting system and extending the solution throughout the space-time plane yields (Sec.~\ref{subsec:appendix:Superluminal_Regime})
    \begin{subequations}\label{eq:Analysis_Superluminal_Solutions}
        \begin{align}
            &\psi_{2}^{-}{\left[\phi_{2}^{-}\right]} = -\frac{\eta_{2}-\eta_{1}}{2\eta_{1}}\frac{1-n_{1}\beta{\left[\left(f_{2}^{-}\right)^{-1}{\left[\phi_{2}^{-}\right]}\right]}}{1+n_{2}\beta{\left[\left(f_{2}^{-}\right)^{-1}{\left[\phi_{2}^{-}\right]}\right]}} \nonumber \\
            &\hspace{3.5cm} \psi_{1}^{+}{\left[f_{1}^{+}{\left[\left(f_{2}^{-}\right)^{-1}{\left[\phi_{2}^{-}\right]}\right]}\right]} \,, \label{eq:Analysis_Superluminal_Solutions_Later-Backward} \\
            &\psi_{2}^{+}{\left[\phi_{2}^{+}\right]} = \frac{\eta_{2} + \eta_{1}}{2\eta_{1}}\frac{1-n_{1}\beta{\left[\left(f_{2}^{+}\right)^{-1}{\left[\phi_{2}^{+}\right]}\right]}}{1-n_{2}\beta{\left[\left(f_{2}^{+}\right)^{-1}{\left[\phi_{2}^{+}\right]}\right]}} \nonumber \\
            &\hspace{3.5cm} \psi_{1}^{+}{\left[f_{1}^{+}{\left[\left(f_{2}^{+}\right)^{-1}{\left[\phi_{2}^{+}\right]}\right]}\right]}\,. \label{eq:Analysis_Superluminal_Solutions_Later-Forward}
        \end{align}
    \end{subequations}
    Notably, the later-backward wave [Eq.~\eqref{eq:Analysis_Superluminal_Solutions_Later-Backward}] depends on $n_{2}$ in the denominator rather than $n_{1}$ as in the subluminal regime [Eq.~\eqref{eq:Analysis_Subluminal_Solutions_Reflection}], reflecting that this wave propagates in the second medium.

\begin{figure*}
    \centering
    \includegraphics[width=1.0\linewidth]{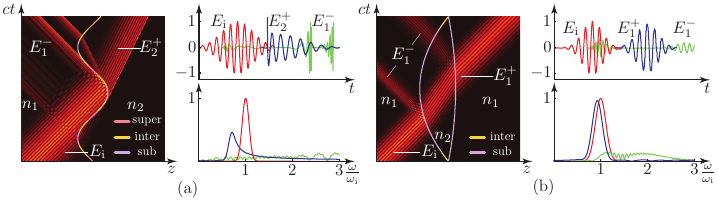} 
    \caption{Illustration examples for the analysis problem [Eqs.~\cref{eq:Analysis_Subluminal_Solutions,eq:Analysis_Superluminal_Solutions,eq:Analysis_Interluminal_Co_Solutions,eq:Analysis_Interluminal_Contra_Solutions}] of electromagnetic scattering at arbitrary accelerated interfaces (Fig.~\ref{fig:Overview}) in different velocity regimes, showing the space-time evolution, time-domain waveforms far from the interface and corresponding spectra. (a)~Single interface with an S-shape trajectory. (b)~Space-time cavity.}
    \label{fig:Examples_Analysis}
\end{figure*}

\begin{figure*}
    \centering
    \includegraphics[width=1.0\linewidth]{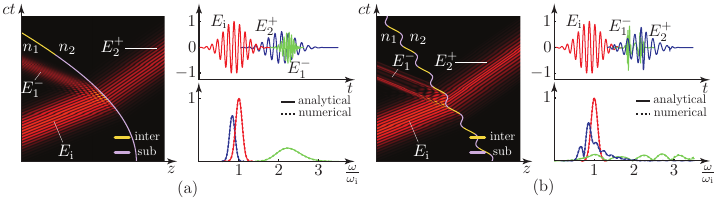} 
    \caption{Full-wave finite-difference time-domain (FDTD) validation of the analysis problem [Eqs.~\cref{eq:Analysis_Subluminal_Solutions,eq:Analysis_Superluminal_Solutions,eq:Analysis_Interluminal_Co_Solutions,eq:Analysis_Interluminal_Contra_Solutions}], showing the space-time evolution, time-domain waveforms far from the interface and corresponding spectra. (a) Constant, proper-accelerated interface. (b) Backwards oscillating trajectory.}
    \label{fig:FDTD_Validation}
\end{figure*}

    \subsubsection{Interluminal}
    \pati{Scattering Solutions}{
    }

    The general boundary conditions for the interluminal regime have not yet been elucidated in the literature. Therefore, we do not derive the interluminal scattering coefficients directly; instead, we adopt the results from~\cite{Deck-Leger2019_Inter_CONF}, which does solve the problem for the particular case of nonmagnetic materials ($\mu_{1} = \mu_{2}$), and we extend these solutions to cover the entire space-time plane (Sec.~\ref{subsec:appendix:Interluminal_Regime}). The scattering behavior in the interluminal regime depends on whether the interface moves codirectionally or contradirectionally and on the incident medium, resulting in different numbers of scattered waves, as illustrated in Fig.~\ref{fig:Overview}. For a forward-propagating incident wave in the first medium and a codirectionally moving interface (Fig.~\ref{fig:Overview}a), the reflected wave is denoted by $\psi_{1}^{-}$ and is given by
    \begin{subequations}\label{eq:Analysis_Interluminal_Co_Solutions}
        \begin{align}
            &\psi_{1}^{-}{\left[\phi_{1}^{-}\right]} = -\left(\frac{1-n_{1}\beta{\left[\left(f_{1}^{-}\right)^{-1}{\left[\phi_{1}^{-}\right]}\right]}}{1+n_{1}\beta{\left[\left(f_{1}^{-}\right)^{-1}{\left[\phi_{1}^{-}\right]}\right]}}\right)^{2} \nonumber \\
            &\hspace{3.5cm} \psi_{1}^{+}{\left[f_{1}^{+}{\left[\left(f_{1}^{-}\right)^{-1}{\left[\phi_{1}^{-}\right]}\right]}\right]} \,.
        \end{align}
    \end{subequations}
    In the contradirectional case (Fig.~\ref{fig:Overview}b), the scattering is more complex, involving a reflected, $\psi_{1}^{-}$, later-backward wave, $\psi_{2}^{-}$, and a transmitted, $\psi_{2}^{+}$, wave, which are given by
    \begin{subequations}\label{eq:Analysis_Interluminal_Contra_Solutions}
        \begin{align}
            \psi_{1}^{-}{\left[\phi_{1}^{-}\right]} &= -\psi_{1}^{+}{\left[f_{1}^{+}{\left[\left(f_{1}^{-}\right)^{-1}{\left[\phi_{1}^{-}\right]}\right]}\right]} \,, \\
            \psi_{2}^{-}{\left[\phi_{2}^{-}\right]} &= -\frac{n_{1}}{n_{2}}\psi_{1}^{+}{\left[f_{1}^{+}{\left[\left(f_{2}^{-}\right)^{-1}{\left[\phi_{2}^{-}\right]}\right]}\right]}\,, \\
            \psi_{2}^{+}{\left[\phi_{2}^{+}\right]} &= \frac{n_{1}}{n_{2}} \psi_{1}^{+}{\left[f_{1}^{+}{\left[\left(f_{2}^{+}\right)^{-1}{\left[\phi_{2}^{+}\right]}\right]}\right]}\,.
        \end{align}
    \end{subequations}

\subsection{Examples}\label{subsec:Examples_Analysis}
    \pati{Annoucement}{
    }
    
    Figure~\ref{fig:Examples_Analysis} shows two representative examples of the analysis problem [Eqs.~\cref{eq:Analysis_Subluminal_Solutions,eq:Analysis_Superluminal_Solutions,eq:Analysis_Interluminal_Co_Solutions,eq:Analysis_Interluminal_Contra_Solutions}], showcasing the space-time evolution, time-domain waveforms and corresponding spectra. Figure~\ref{fig:Examples_Analysis}a shows an S-shaped trajectory while Fig.~\ref{fig:Examples_Analysis}b demonstrates an example of a space-time cavity. Both examples use nonmagnetic materials to ensure the validity of the adopted interluminal solutions [Eq.~\eqref{eq:Analysis_Interluminal_Co_Solutions} and Eqs.~\eqref{eq:Analysis_Interluminal_Contra_Solutions}].

    \pati{S-Shape}{
    }

    In the S-shaped trajectory example (Fig.~\ref{fig:Examples_Analysis}a), the interface smoothly oscillates through subluminal, interluminal and superluminal velocity regimes, producing the scattering behavior described in Sec.~\ref{sec:Physics_of_Accelerated_Interfaces}. The motion imprints a clear chirping pattern on the scattered waves, visible in the time-domain signal of the transmitted wave. Moreover, the interface causes multiple scattering events. For example, the center part of the incoming pulse scatters at a contramoving interluminal section of the interface, resulting in a back-scattered wave into the second medium. As the interface reverses direction, it interacts again with this previously scattered wave, producing additional scattering.

    \pati{Space-Time Cavity}{
    }

    The solutions presented in this section apply not only to a single interface but also to configurations involving multiple interfaces, since each scattering event may be treated as independent of prior interactions. Figure~\ref{fig:Examples_Analysis}b shows an example of a space-time cavity formed by two moving interfaces enclosing a region of different refractive index for a finite time. Such a cavity may act as a dynamic analogue of a Fabry–Pérot resonator: it opens and closes in both space and time, trapping waves through repeated reflections. As the wave encounters the first interface, part of it reflects with a nonuniform Doppler shift, determined by Eq.~\eqref{eq:Doppler_Shift}. Inside the cavity, the wave undergoes multiple scattering events, each introducing additional nonuniform Doppler shifts.

    \pati{FDTD Validation}{
    }

    To validate our analytical results, we have performed full-wave finite-difference time-domain (FDTD) simulations using a generalized Yee grid~\cite{Deck-Leger2022_FDTD_PUB,Bahrami2023_FDTD_PUB} that ensures the enforcement of proper moving boundary conditions. The simulation results are shown in Fig.~\ref{fig:FDTD_Validation}. They include a constant, proper-accelerated interface (Fig.~\ref{fig:FDTD_Validation}a) and an oscillatory backward moving interface (Fig.~\ref{fig:FDTD_Validation}b). Excellent agreement with the analytical predictions is observed. Since the theory is exact, any discrepancies are attributable to numerical approximations, as we verified by increasing the computational mesh density within the memory limits of our computer system. These examples are restricted to subluminal and interluminal velocities, because the used FDTD code tends to be unstable in the superluminal regime.

\section{Synthesis}\label{sec:Synthesis}
    \begin{figure*}
        \centering
        \includegraphics[width=1.0\linewidth]{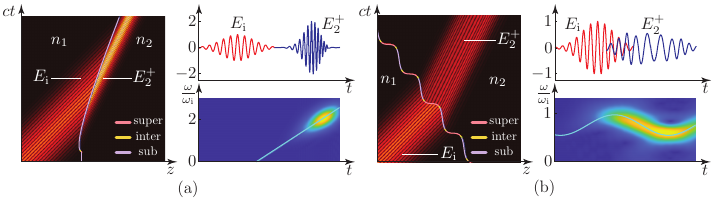} 
        \caption{Illustrative examples of the synthesis problem [Eqs.~\eqref{eq:Synthesis_Full_Trajectory}] illustrated through space-time diagrams, time-domain scattered waveforms far from the interface and corresponding spectrograms. (a)~Linear chirp. (b)~Oscillatory chirp.}
        \label{fig:Example_Synthesis}
    \end{figure*}

    \subsection{Interface Equation of Motion}\label{subsec:Interface_Equation_of_Motion}
        \pati{Problem Statement}{
        }

        In Sec.~\ref{sec:Analysis}, we showed that the scattered waves exhibit a specific temporal chirping profile governed by the acceleration of the interface [Eq.~\eqref{eq:Doppler_Shift}]. In this section, we address the inverse (synthesis) problem: given a desired chirping profile, denoted by $\varphi{\left[t\right]}$, for the \emph{transmitted} wave, we seek the interface trajectory, $z{\left[t\right]}$, that produces it. Although the synthesis focuses on the transmitted component, the methodology is equally applicable to other scattered waves, including reflected and later-backward propagating components.

        \pati{Interface Trajectory Solution}{
        }

        Solving this inverse problem is nontrivial because the phase argument of the transmitted wave [Eq.~\eqref{eq:Analysis_Subluminal_Solutions_Transmission}] contains nested dependencies of the unknown trajectory, $z{\left[t\right]}$: $f_{1}^{+}$ is evaluated at the inverse of $f_{2}^{+}$, and both depend on $z{\left[t\right]}$ [Eq.~\eqref{eq:General_Definition_Function_f}].
        Nevertheless, by introducing a suitable change of variables, we obtain an explicit expression for the interface equation of motion (Sec.~\ref{subsec:appendix:Interface_Profile}), viz.,
        \begin{subequations}\label{eq:Synthesis_Full_Trajectory}
            \begin{align}\label{eq:Synthesis_Trajectory}
                z{\left[t\right]} = \frac{u_{1}u_{2}}{u_{1}-u_{2}}\left(x{\left[t\right]} - \varphi{\left[x{\left[t\right]}\right]}\right)\,,
            \end{align}
            where the auxiliary function $x{\left[t\right]}$ satisfies
            \begin{align}\label{eq:Synthesis_Transformation}
                \frac{u_{2}x - u_{1}\varphi{\left[x\right]}}{u_{1}-u_{2}} = t\,,
            \end{align}
        \end{subequations}
        For complex chirping profiles, Eq.~\eqref{eq:Synthesis_Transformation} may need to be solved numerically.

        \pati{Uniform Doppler Shift Example}{
        }
        
        To illustrate Eqs.~\eqref{eq:Synthesis_Full_Trajectory}, consider the simple case of a uniform Doppler shift, viz., $\varphi{\left[t\right]} = at$. Solving Eq.~\eqref{eq:Synthesis_Transformation} for $x{\left[t\right]}$ yields $x{\left[t\right]} = \left(u_{1}-u_{2}\right)/\left(u_{2}-au_{1}\right)t$, which can then be substituted in Eq.~\eqref{eq:Synthesis_Trajectory}, giving
        \begin{equation}\label{eq:Synthesis_Example_Uniform_Doppler_Interface}
            z{\left[t\right]} = \frac{1-a}{n_{1}-an_{2}}ct\,.
        \end{equation}
        Equation~\eqref{eq:Synthesis_Example_Uniform_Doppler_Interface} describes an interface moving at a constant velocity $\beta_{0} = \left(1-a\right)/\left(n_{1}-an_{2}\right)$, resulting in the Doppler shift $\left(1-n_{1}\beta_{0}\right)/\left(1-n_{2}\beta_{0}\right) = a$ [Eq.~\eqref{eq:Doppler_Shift}], as expected.

        \pati{Limitation Arbitrary Chirping}{
        }
        
        Although Eqs.~\eqref{eq:Synthesis_Full_Trajectory} permit a priori arbitrary chirping profiles, they include a fundamental constraint arising from the physics of the scattering process: not every configuration supports a transmitted wave in the second medium. Specifically, for a forward incident wave, codirectional interluminal and superluminal interfaces do not generate a transmitted component in the second medium (Fig.~\ref{fig:Overview}). Therefore, we must restrict the interface velocity to remain below the lower codirectional interluminal bound, viz., $\beta{\left[t\right]} < 1/n_{2}$. This requirement imposes the following condition on the admissible chirping profiles (Sec.~\ref{subsec:appendix:Limitation})
        \begin{equation}\label{eq:Arbitray_Chirping_Constraint}
            \dot{\varphi}{\left[x{\left[t\right]}\right]} > \frac{n_{1}}{n_{2}}\,,
        \end{equation}
        where the dot denotes differentiation with respect to time, indicating that the instantaneous frequency must exceed this threshold for consistent transmission.

    \subsection{Examples}\label{subsec:Examples_Synthesis}
        \pati{Announcement}{
        }
        
        Figure~\ref{fig:Example_Synthesis} demonstrates the synthesis problem [Eqs.~\eqref{eq:Synthesis_Full_Trajectory}] for two chirping profiles---a linear chirp (Fig.~\ref{fig:Example_Synthesis}a) and an oscillatory chirp (Fig.~\ref{fig:Example_Synthesis}b), both satisfying the chirping constraint [Eq.~\eqref{eq:Arbitray_Chirping_Constraint}]. Each case is illustrated through a space-time diagram, time domain waveforms and related spectrograms. The red curve in each spectrogram marks the analytically prescribed chirping profile. Both examples assume impedance matching ($\eta_{1} = \eta_{2}$) to suppress spurious reflections and focus solely on the desired transmitted wave.

        \pati{Linear Chirp}{
        }

        Figure~\ref{fig:Example_Synthesis}a shows the synthesis problem for a linear chirping profile, viz., $\varphi\left[t\right] = at + bt^{2}$, with $a$ the transmitted Doppler shift and $b$ the chirp rate.
        After a stationary period, the interface moves for a very short time at an extremely high negative velocity (flat segment) and starts accelerating toward the interluminal asymptote $\beta = 1/n_{2}$, where the local Doppler shift diverges [Eq.~\eqref{eq:Doppler_Shift}]. Consequently, the transmitted frequency continuously upshifts over time, which is clearly visible in both the time-domain signal and its spectrogram.

        \pati{Oscillating Chirp}{
        }
        
        Figure~\ref{fig:Example_Synthesis}b presents a more complex situation, where the chirping profile combines a linear shift with a sinusoidal modulation, $\varphi{\left[t\right]} = at + A\sin\left(2\pi \omega_{\text{c}} t\right)$, where $A$ is the amplitude and $\omega_{\text{c}}$ the oscillation frequency  of the chirp profile. The synthesized interface trajectory consists of a backward drift with superimposed oscillations, causing it to traverse multiple velocity regimes. Physically, this motion can be viewed as the combination of a steady backward velocity (responsible for $a$) and an oscillatory component that modulates the instantaneous frequency, producing the characteristic sinusoidal chirp pattern evident in the spectrogram.

\section{Conclusions and Discussion}\label{sec:Conclusions}
    \pati{Conclusions}{
    }

    We have presented exact field solutions for scattering at an arbitrarily accelerated interface, covering the three interface velocity regimes. Unlike previous approaches that rely on the frame hopping method, we solved the problem directly in the laboratory frame through a suitable change of variables. Our results show that accelerating interfaces introduce phenomena that do not occur in uniformly moving interfaces, including transitions between multiple velocity regimes, multiple scattering events and generalized frequency chirping. In addition, we have addressed the inverse problem, demonstrating how tailored acceleration can be used to realize prescribed chirping profiles. Solving this synthesis problem is a crucial step toward designing space-time interfaces capable of producing targeted frequency modulation, with potential applications in analog signal processing~\cite{Caloz2013_PUB} and chirped pulse shaping~\cite{Strickland1985_CPA_PUB}.

    \pati{Experimental Outlook}{
    }
    
    Experimental realization of accelerated interfaces, while challenging, appears feasible based on existing demonstrations of purely temporal interfaces in both microwave~\cite{Alu2023_Temp_Refl_PUB,Alu2023_Coherent_Wave_Control_PUB,Peroulis2024_Time_Ref_TEM_PUB} and optical~\cite{Boyd2020_Time_Refr_ENZ_TEM_PUB,Segev2023_Single_Cycle_TEM_PUB,Kinsey2025_ST_Knife_TEM_PUB} systems. In the microwave regime, the effective impedance of artificial transmission lines can be modulated through high-speed electronic switches driven in a controlled, accelerated sequence via synchronized timing or corporate feed networks with optical delay lines. In the optical regime, pump-probe setups are able to generate temporal refractive-index fronts. By positioning the pump beam oblique with respect to the substrate and before a focusing element, the resulting curved pump wavefront excites the substrate at varying incidence angles, thereby producing an effectively accelerated interface. These approaches indicate that accelerated space–time interfaces could be experimentally investigated in both microwave and optical domains.
    
    \noindent\textbf{Author contributions} \\
    K.D.K. performed the bulk of the work. A.B. assisted with the calculations. C.C. supervised the work. All authors accepted responsibility for the entire content of the manuscript and approved its submission.

    \noindent\textbf{Competing interests} \\
    The authors declare that they have no competing interests.

    \noindent\textbf{Acknowledgments} \\
    K.D.K. is supported by the Research Foundation -- Flanders (FWO) doctoral fellowship 1174526N.

\appendix
\section{Scattering at Accelerated Interfaces}\label{sec:appendix:Scattering_at_Accelerated_Interfaces}
    \pati{Announcement}{ 
    }
    
    This section derives the electromagnetic scattering solutions at a nonuniformly (accelerating) moving interface. The interface separates two isotropic, linear and nondispersive media, characterized by refractive indices $n_{1,2}$ and impedances $\eta_{1,2}$. The interface equation of motion has the explicit form $z{\left[t\right]}$, with normalized velocity $\beta{[t]} = \dd{z}/\dd{(ct)}$. We consider a one-dimensional geometry: the interface moves along the $z$-direction, the electric field is polarized along the $x$-direction and magnetic field lies along the $y$-direction. A general traveling wave in each medium takes the form
    \begin{subequations}\label{eq:appendix:General_Waveforms}     
        \begin{align}
            E_{i}^{\pm} &= \psi_{i}^{\pm}{\left[\phi_{i}^{\pm}\right]}\,, 
            &H_{i}^{\pm} &= \pm \frac{1}{\eta_{i}}\psi_{i}^{\pm}{\left[\phi_{i}^{\pm}\right]}\,, \\
            D_{i}^{\pm} &= \frac{1}{u_{i}\eta_{i}}\psi_{i}^{\pm}{\left[\phi_{i}^{\pm}\right]}\,, 
            &B_{i}^{\pm} &= \pm \frac{1}{u_{i}}\psi_{i}^{\pm}{\left[\phi_{i}^{\pm}\right]}\,, 
        \end{align}
    \end{subequations}
    where we have used square brackets, $\left[\cdot\right]$, to indicate the arguments of the wave functions, $\psi_{i}^{\pm}$, and
    \begin{equation}\label{eq:appendix:Traveling_Wave_Variables}
        \phi_{i}^{\pm} = z/u_{i} \mp t
    \end{equation}
    are the traveling wave variables. The subscript, $i = 1,2$, denotes the medium, while the superscript, $\pm$, indicates forward ($+$) or backward ($-$) propagating waves. The wave speed in medium $i$ is given by $u_{i} = c/n_{i}$.

    \subsection{Subluminal Regime}\label{subsec:appendix:Subluminal_Regime}
    \pati{Subluminal Regime}{
    }
    
    In the subluminal velocity regime ($\left|\beta{\left[t\right]}\right| < 1/n_{2}$), scattering consists of a reflected wave, in the same medium but opposite direction as the incoming wave, and a transmitted wave, in the opposite medium, propagating in the same direction as the incident wave. The boundary conditions are the continuity of $\boldsymbol{E} + c\boldsymbol{\beta} \times \boldsymbol{B}$ and $\boldsymbol{H} - c\boldsymbol{\beta} \times \boldsymbol{D}$ at the interface~\cite{Caloz2019b_ST_Metamaterials_PUB}. Considering a forward-propagating incident wave in the first medium, $\psi_{1}^{+}$, the reflected and transmitted waves are $\psi_{1}^{-}$ and $\psi_{2}^{+}$, respectively. The boundary conditions then become
    \begin{subequations}\label{eq:appendix:Subluminal_Boundary_Conditions}
        \begin{align}
            &\left(E_{1}^{+} - c\beta B_{1}^{+}\right) + \left(E_{1}^{-} - c\beta B_{1}^{-}\right)\bigg|_{z=z{\left[t^{\star}\right]}} \nonumber \\
            &\hspace{3.5cm}= \left(E_{2}^{+} - c\beta B_{2}^{+}\right)\bigg|_{z=z{\left[t^{\star}\right]}} \,, \\
            &\left(H_{1}^{+} - c\beta D_{1}^{+}\right) + \left(H_{1}^{-} - c\beta D_{1}^{-}\right)\bigg|_{z=z{\left[t^{\star}\right]}} \nonumber \\
            &\hspace{3.4cm}= \left(H_{2}^{+} - c\beta D_{2}^{+}\right)\bigg|_{z=z{\left[t^{\star}\right]}} \,,
        \end{align}
    \end{subequations}
    where $t^{\star}$ represents the \emph{scattering time} at the interface. Inserting the general waveforms [Eqs.~\eqref{eq:appendix:General_Waveforms}] into these boundary conditions [Eqs.~\eqref{eq:appendix:Subluminal_Boundary_Conditions}] and evaluating the result at the interface, yields the following system of equations for $\psi_{1}^{-}$ and $\psi_{2}^{+}$:
    \begin{widetext}
        \begin{subequations}\label{eq:appendix:Subluminal_System_of_Equations}
            \begin{align}
                \left(1-n_{1}\beta{\left[t^{\star}\right]}\right)\psi_{1}^{+}{\left[\frac{z{\left[t^{\star}\right]}}{u_{1}}-t^{\star}\right]} + \left(1+n_{1}\beta{\left[t^{\star}\right]}\right)\psi_{1}^{-}{\left[\frac{z{\left[t^{\star}\right]}}{u_{1}}+t^{\star}\right]} &= \left(1-n_{2}\beta{\left[t^{\star}\right]}\right)\psi_{2}^{+}{\left[\frac{z{\left[t^{\star}\right]}}{u_{2}}-t^{\star}\right]}\,, \\
                \frac{1}{\eta_{1}}\left(1-n_{1}\beta{\left[t^{\star}\right]}\right)\psi_{1}^{+}{\left[\frac{z{\left[t^{\star}\right]}}{u_{1}}-t^{\star}\right]} - \frac{1}{\eta_{1}}\left(1+n_{1}\beta{\left[t^{\star}\right]}\right)\psi_{1}^{-}{\left[\frac{z{\left[t^{\star}\right]}}{u_{1}}+t^{\star}\right]} &= \frac{1}{\eta_{2}}\left(1-n_{2}\beta{\left[t^{\star}\right]}\right)\psi_{2}^{+}{\left[\frac{z{\left[t^{\star}\right]}}{u_{2}}-t^{\star}\right]}\,.
            \end{align}
        \end{subequations}
    \end{widetext}
    Solving Eqs.~\eqref{eq:appendix:Subluminal_System_of_Equations} for $\psi_{1}^{-}$ and $\psi_{2}^{+}$, we obtain
    \begin{subequations}\label{eqa:p1m_p2p}
        \begin{align}\label{eq:appendix:Subluminal_Solution_System_of_Equations_With_Function_f}
            \psi_{1}^{-}{\left[f_{1}^{-}{\left[t^{\star}\right]}\right]} &= \frac{\eta_{2}-\eta_{1}}{\eta_{2} + \eta_{1}}\frac{1-n_{1}\beta{\left[t^{\star}\right]}}{1+n_{1}\beta{\left[t^{\star}\right]}}\psi_{1}^{+}{\left[f_{1}^{+}{\left[t^{\star}\right]}\right]}\,, \\
            \psi_{2}^{+}{\left[f_{2}^{+}{\left[t^{\star}\right]}\right]} &= \frac{2\eta_{2}}{\eta_{2} + \eta_{1}}\frac{1-n_{1}\beta{\left[t^{\star}\right]}}{1-n_{2}\beta{\left[t^{\star}\right]}}\psi_{1}^{+}{\left[f_{1}^{+}{\left[t^{\star}\right]}\right]}\,,
        \end{align}
    \end{subequations}
    where we used the shortcut function $f_{i}^{\pm}$, defined as
    \begin{equation}\label{eq:appendix:Subluminal_General_Definition_Function_f}
        f_{i}^{\pm}{\left[t^{\star}\right]} = \frac{z{\left[t^{\star}\right]}}{u_{i}} \mp t^{\star}\,.
    \end{equation}
    Equations~\eqref{eqa:p1m_p2p} are valid only at the interface, since the time parameter they involve is the scattering time at the interface. To determine the scattered waves at \emph{any space-time point}, we require that the arguments on the left-hand side of Eqs.~\eqref{eqa:p1m_p2p} match the traveling wave variables, $\phi_{1}^{-}$ and $\phi_{2}^{+}$:
    \begin{subequations}\label{eq:appendix:Subluminal_Desired_Traveling_Waveform}
        \begin{align}
            f_{1}^{-}{\left[t^{\star}\right]} &= \phi_{1}^{-} \,, \\
            f_{2}^{+}{\left[t^{\star}\right]} &= \phi_{2}^{+} \,.
        \end{align}
    \end{subequations}
    Next, we invert both sides in Eqs.~\eqref{eq:appendix:Subluminal_Desired_Traveling_Waveform}, yielding the changes of variables
    \begin{subequations}\label{eq:appendix:Subluminal_Change_of_Variables}
        \begin{align}
            t^{\star} &\mapsto \left(f_{1}^{-}\right)^{-1}{\left[\phi_{1}^{-}\right]} \,, \label{eq:appendix:Subluminal_Change_of_Variables_Transmission}\\
            t^{\star} &\mapsto \left(f_{2}^{+}\right)^{-1}{\left[\phi_{2}^{+}\right]} \,.
        \end{align}
    \end{subequations}
    Substituting Eqs.~\eqref{eq:appendix:Subluminal_Change_of_Variables} into Eqs.~\eqref{eqa:p1m_p2p} yields the final result:
    \begin{subequations}\label{eq:appendix:Subluminal_Scattering_Solutions}
        \begin{align}
            &\psi_{1}^{-}{\left[\phi_{1}^{-}\right]} = \frac{\eta_{2}-\eta_{1}}{\eta_{2}+\eta_{1}}\frac{1-n_{1}\beta{\left[\left(f_{1}^{-}\right)^{-1}{\left[\phi_{1}^{-}\right]}\right]}}{1+n_{1}\beta{\left[\left(f_{1}^{-}\right)^{-1}{\left[\phi_{1}^{-}\right]}\right]}} \nonumber \\
            &\hspace{3.5cm}\psi_{1}^{+}{\left[f_{1}^{+}{\left[\left(f_{1}^{-}\right)^{-1}{\left[\phi_{1}^{-}\right]}\right]}\right]} \,, \\  
            &\psi_{2}^{+}{\left[\phi_{2}^{+}\right]} = \frac{2\eta_{2}}{\eta_{2}+\eta_{1}}\frac{1-n_{1}\beta{\left[\left(f_{2}^{+}\right)^{-1}{\left[\phi_{2}^{+}\right]}\right]}}{1-n_{2}\beta{\left[\left(f_{2}^{+}\right)^{-1}{\left[\phi_{2}^{+}\right]}\right]}} \nonumber \\
            &\hspace{3.5cm}\psi_{1}^{+}{\left[f_{1}^{+}{\left[\left(f_{2}^{+}\right)^{-1}{\left[\phi_{2}^{+}\right]}\right]}\right]}\,, \label{eq:appendix:Subluminal_Scattering_Solutions_Transmitted_Wave}
        \end{align}
    \end{subequations}
    where the $t$'s within $\phi_i^{\pm}$ [Eq.~\eqref{eq:appendix:Traveling_Wave_Variables}] represent now the \emph{general laboratory time}, applying both to scattering and propagation times.

    \pati{Chirping}{ 
    }
    
    The Doppler-induced frequency shift may be obtained by differentiating the phase argument of the scattered waves in Eqs.~\eqref{eq:appendix:Subluminal_Scattering_Solutions} with respect to time. Since both scattered waves have a phase of the form $f_{1}^{+}{\left[t^{\star}\right]}$, with $t^{\star} = \left(f_{i}^{\pm}\right)^{-1}{\left[\phi_{i}^{\pm}\right]}$ [Eq.~\eqref{eq:appendix:Subluminal_Change_of_Variables}], we find, using the definition of $f_{1}^{+}$ in Eq.~\eqref{eq:appendix:Subluminal_General_Definition_Function_f}:
    \begin{align}
            \omega_{i}^{\pm}{\left[t^{\star}\right]} &= -\frac{\partial}{\partial t}\left(f_{1}^{+}{\left[t^{\star}\right]}\right) \nonumber \\
            &= \left(1- \frac{1}{u_{1}}\frac{\dd{z{\left[t^{\star}\right]}}}{\dd{t^{\star}}}\right)\frac{\partial t^{\star}}{\partial t} \nonumber \\
            &= \left(1- n_{1}\beta{\left[t^{\star}\right]}\right)\frac{\partial t^{\star}}{\partial t} \,, \label{eq:appendix:Doppler_Shift_First_Result_Doppler}
    \end{align}
    where $\beta{\left[t^{\star}\right]} = \dd{z{\left[t^{\star}\right]}}/\dd{{\left(ct^{\star}\right)}}$ and $u_{1} = c/n_{1}$ were used in the last equality. Using the identity $\partial_{x}\left(g^{-1}{\left[x\right]}\right) = 1/g'{\left[g^{-1}{\left[x\right]}\right]}$, we find for the last factor in Eq.~\eqref{eq:appendix:Doppler_Shift_First_Result_Doppler}:
    \begin{align}
            \frac{\partial t^{\star}}{\partial t} &= \frac{\partial}{\partial t}\left(\left(f_{i}^{\pm}\right)^{-1}{\left[\phi_{i}^{\pm}\right]}\right) \nonumber \\
            &= \mp\frac{1}{\partial_{t^{\star}}\left(f_{i}^{\pm}{\left[t^{\star}\right]}\right)} \nonumber \\
            &= \mp\frac{1}{n_{i}\beta{\left[t^{\star}\right]} \mp 1}\,. \label{eq:appendix:Doppler_Shift_Intermediate_Result_Partial_Derivative_tstar}
    \end{align}
    Inserting Eq.~\eqref{eq:appendix:Doppler_Shift_Intermediate_Result_Partial_Derivative_tstar} into Eq.~\eqref{eq:appendix:Doppler_Shift_First_Result_Doppler} yields the frequency shift
    \begin{equation}\label{eq:appendix:Doppler_Shift}
        \omega_{i}^{\pm}{\left[t^{\star}\right]} = \pm\frac{n_{1}\beta{\left[t^{\star}\right]} - 1}{n_{i}\beta{\left[t^{\star}\right]} \mp 1}\,,
    \end{equation}
    where $t^{\star} = \left(f_{i}^{\pm}\right)^{-1}{\left[\phi_{i}^{\pm}\right]}$ [Eq.~\eqref{eq:appendix:Subluminal_Change_of_Variables}]. If $\beta{\left[t^{\star}\right]}$ is a non-constant function, Eq.~\eqref{eq:appendix:Doppler_Shift} describes a time-varying Doppler shift---a chirp---whose profile is determined by the velocity trajectory.

    \subsection{Superluminal Regime}\label{subsec:appendix:Superluminal_Regime}
    \pati{Superluminal Regime}{ 
    }
    
    In the superluminal regime ($\left|\beta{\left[t\right]}\right| > 1/n_{1}$), scattering consists of a later-backward wave, in the opposite medium and opposite direction as the incoming wave, and a transmitted wave, in the opposite medium, propagating in the same direction as the incident wave. The boundary conditions are again the continuity of $\boldsymbol{E} + c\boldsymbol{\beta} \times \boldsymbol{B}$ and $\boldsymbol{H} - c\boldsymbol{\beta} \times \boldsymbol{D}$ at the interface~\cite{Caloz2019b_ST_Metamaterials_PUB}. Considering a forward-propagating incident wave in the first medium, $\psi_{1}^{+}$, the later-backward and transmitted waves are $\psi_{2}^{-}$ and $\psi_{2}^{+}$, respectively. The boundary conditions then become:
    \begin{subequations}\label{eq:appendix:Superluminal_Boundary_Conditions}
        \begin{align}
            &\left(E_{1}^{+} - c\beta B_{1}^{+}\right)\bigg|_{z=z{\left[t^{\star}\right]}} = \left(E_{2}^{-} - c\beta B_{2}^{-}\right) \nonumber \\
            &\hspace{3.3cm}+ \left(E_{2}^{+} - c\beta B_{2}^{+}\right)\bigg|_{z=z{\left[t^{\star}\right]}} \,, \\
            &\left(H_{1}^{+} - c\beta D_{1}^{+}\right)\bigg|_{z=z{\left[t^{\star}\right]}} = \left(H_{2}^{-} - c\beta D_{2}^{-}\right) \nonumber \\
            &\hspace{3.3cm} + \left(H_{2}^{+} - c\beta D_{2}^{+}\right)\bigg|_{z=z{\left[t^{\star}\right]}} \,,
        \end{align}
    \end{subequations}
    where once again $t^{\star}$ represents the scattering time at the interface. Inserting the general waveforms [Eqs.~\eqref{eq:appendix:General_Waveforms}] into these boundary conditions [Eqs.~\eqref{eq:appendix:Superluminal_Boundary_Conditions}] and evaluating the result at the interface gives the following system of equations:
    \begin{widetext}
        \begin{subequations}\label{eq:appendix:Superluminal_System_of_Equations}
            \begin{align}
                \left(1-n_{1}\beta{\left[t^{\star}\right]}\right)\psi_{1}^{+}{\left[\frac{z{\left[t^{\star}\right]}}{u_{1}}-t^{\star}\right]} &= \left(1+n_{2}\beta{\left[t^{\star}\right]}\right)\psi_{2}^{-}{\left[\frac{z{\left[t^{\star}\right]}}{u_{2}}+t^{\star}\right]} + \left(1-n_{2}\beta{\left[t^{\star}\right]}\right)\psi_{2}^{+}{\left[\frac{z{\left[t^{\star}\right]}}{u_{2}}-t^{\star}\right]} \,, \\
                \frac{1}{\eta_{1}}\left(1-n_{1}\beta{\left[t^{\star}\right]}\right)\psi_{1}^{+}{\left[\frac{z{\left[t^{\star}\right]}}{u_{1}}-t^{\star}\right]} &= -\frac{1}{\eta_{2}}\left(1+n_{2}\beta{\left[t^{\star}\right]}\right)\psi_{2}^{-}{\left[\frac{z{\left[t^{\star}\right]}}{u_{2}}+t^{\star}\right]} + \frac{1}{\eta_{2}}\left(1-n_{2}\beta{\left[t^{\star}\right]}\right)\psi_{2}^{+}{\left[\frac{z{\left[t^{\star}\right]}}{u_{2}}-t^{\star}\right]}\,.
            \end{align}
        \end{subequations}
    \end{widetext}
    Solving Eqs.~\eqref{eq:appendix:Superluminal_System_of_Equations} for the scattered waves yields
    \begin{subequations}\label{eq:appendix:Superluminal_Solution_System_of_Equations_With_Function_f}
        \begin{align}
            \psi_{2}^{-}{\left[f_{2}^{-}{\left[t^{\star}\right]}\right]} &= -\frac{\eta_{2}-\eta_{1}}{2\eta_{1}}\frac{1-n_{1}\beta{\left[t^{\star}\right]}}{1+n_{2}\beta{\left[t^{\star}\right]}}\psi_{1}^{+}{\left[f_{1}^{+}{\left[t^{\star}\right]}\right]}\,, \\
            \psi_{2}^{+}{\left[f_{2}^{+}{\left[t^{\star}\right]}\right]} &= \frac{\eta_{2} + \eta_{1}}{2\eta_{1}}\frac{1-n_{1}\beta{\left[t^{\star}\right]}}{1-n_{2}\beta{\left[t^{\star}\right]}}\psi_{1}^{+}{\left[f_{1}^{+}{\left[t^{\star}\right]}\right]}\,,
        \end{align}
    \end{subequations}
    where the functions $f_{i}^{\pm}$ are defined in Eq.~\eqref{eq:appendix:Subluminal_General_Definition_Function_f}. Again, equations~\eqref{eq:appendix:Superluminal_Solution_System_of_Equations_With_Function_f} are valid only at the interface, and to extend them across the space-time domain, we enforce that the arguments in the left hand side of Eqs.~\eqref{eq:appendix:Superluminal_Solution_System_of_Equations_With_Function_f} take the traveling waveform variables, $\phi_{2}^{-}$ and $\phi_{2}^{+}$:
    \begin{subequations}\label{eq:appendix:Superluminal_Desired_Traveling_Waveform_With_Function}
        \begin{align}
            f_{2}^{-}{\left[t^{\star}\right]} &= \phi_{2}^{-}\,, \\
            f_{2}^{+}{\left[t^{\star}\right]} &= \phi_{2}^{+}\,.
        \end{align}
    \end{subequations}
    Inverting both sides of Eqs.~\eqref{eq:appendix:Superluminal_Desired_Traveling_Waveform_With_Function} yields the change of variables
    \begin{subequations}\label{eq:appendix:Superluminal_Change_of_Variables}
        \begin{align}
            t^{\star} &\mapsto \left(f_{2}^{-}\right)^{-1}{\left[\phi_{2}^{-}\right]} \,, \\
            t^{\star} &\mapsto \left(f_{2}^{+}\right)^{-1}{\left[\phi_{2}^{+}\right]} \,.
        \end{align}
    \end{subequations}
    Substituting Eqs.~\eqref{eq:appendix:Superluminal_Change_of_Variables} into Eqs.~\eqref{eq:appendix:Superluminal_Solution_System_of_Equations_With_Function_f} yields the final result
    \begin{subequations}\label{eq:appendix:Superluminal_Scattering_Solutions}
        \begin{align}
        &\psi_{2}^{-}{\left[\phi_{2}^{-}\right]} = -\frac{\eta_{2}-\eta_{1}}{2\eta_{1}}\frac{1-n_{1}\beta{\left[\left(f_{2}^{-}\right)^{-1}{\left[\phi_{2}^{-}\right]}\right]}}{1+n_{2}\beta{\left[\left(f_{2}^{-}\right)^{-1}{\left[\phi_{2}^{-}\right]}\right]}} \nonumber \\
        &\hspace{3.5cm}\psi_{1}^{+}{\left[f_{1}^{+}{\left[\left(f_{2}^{-}\right)^{-1}{\left[\phi_{2}^{-}\right]}\right]}\right]} \,, \\  
        &\psi_{2}^{+}{\left[\phi_{2}^{+}\right]} = \frac{\eta_{2} + \eta_{1}}{2\eta_{1}}\frac{1-n_{1}\beta{\left[\left(f_{2}^{+}\right)^{-1}{\left[\phi_{2}^{+}\right]}\right]}}{1-n_{2}\beta{\left[\left(f_{2}^{+}\right)^{-1}{\left[\phi_{2}^{+}\right]}\right]}} \nonumber \\
        &\hspace{3.5cm}\psi_{1}^{+}{\left[f_{1}^{+}{\left[\left(f_{2}^{+}\right)^{-1}{\left[\phi_{2}^{+}\right]}\right]}\right]}\,.
        \end{align}
    \end{subequations}
    The Doppler frequency shift can be determined in a similar fashion to the previous section.

    \subsection{Interluminal Regime}\label{subsec:appendix:Interluminal_Regime}
    \pati{Interluminal Regime}{ 
    }
    
    The boundary conditions for the interluminal regime ($1/n_{2} \leq \left|\beta{\left[t\right]}\right| \leq 1/n_{1}$) are unknown and hence do not admit a general waveform solution, not even for uniformly moving interfaces. However, if the media are nonmagnetic ($\mu_{1} = \mu_{2}$), the problem becomes solvable~\cite{Ostrovskii1967_Inter_PUB,Deck-Leger2019_Inter_CONF}. Under this assumption, we apply the solutions from~\cite{Deck-Leger2019_Inter_CONF} across space-time using a change of variables. As before, we consider a forward-propagating incident wave in the first medium, $\psi_{1}^{+}$. In the \emph{comoving} case (Fig.~\ref{fig:Overview}a), there is only reflection, $\psi_{1}^{-}$, with scattering coefficient at the interface $\left(1-n_{1}\beta{\left[t^{\star}\right]}\right)^{2}/\left(1+n_{1}\beta{\left[t^{\star}\right]}\right)^{2}$. Applying the change of variables as in the subluminal case for reflection [Eq.~\eqref{eq:appendix:Subluminal_Change_of_Variables}] yields
    \begin{align}
        &\psi_{1}^{-}{\left[\phi_{1}^{-}\right]} = -\left(\frac{1-n_{1}\beta{\left[\left(f_{1}^{-}\right)^{-1}{\left[\phi_{1}^{-}\right]}\right]}}{1+n_{1}\beta{\left[\left(f_{1}^{-}\right)^{-1}{\left[\phi_{1}^{-}\right]}\right]}}\right)^{2}  \nonumber \\
        &\hspace{3.5cm}\psi_{1}^{+}{\left[f_{1}^{+}{\left[\left(f_{1}^{-}\right)^{-1}{\left[\phi_{1}^{-}\right]}\right]}\right]} \,. \label{eq:appendix:Interluminal_Co_Scattering_Solutions}
    \end{align}
    For the \emph{contramoving} case (Fig.~\ref{fig:Overview}b), there are three scattering coefficients: $-1$ for reflection ($\psi_{1}^{-}$), $-n_{1}/n_{2}$ for later-backward ($\psi_{2}^{-}$) and $n_{1}/n_{2}$ for transmission ($\psi_{2}^{+}$). Applying the change of variables to these waves results in the following final solutions:
    \begin{subequations}\label{eq:appendix:Interluminal_Contra_Scattering_Solutions}
        \begin{align}
            \psi_{1}^{-}{\left[\phi_{1}^{-}\right]} &= -\psi_{1}^{+}{\left[f_{1}^{+}{\left[\left(f_{1}^{-}\right)^{-1}{\left[\phi_{1}^{-}\right]}\right]}\right]} \,, \\
            \psi_{2}^{-}{\left[\phi_{2}^{-}\right]} &= -\frac{n_{1}}{n_{2}}\psi_{1}^{+}{\left[f_{1}^{+}{\left[\left(f_{2}^{-}\right)^{-1}{\left[\phi_{2}^{-}\right]}\right]}\right]}\,, \\
            \psi_{2}^{+}{\left[\phi_{2}^{+}\right]} &= \frac{n_{1}}{n_{2}} \psi_{1}^{+}{\left[f_{1}^{+}{\left[\left(f_{2}^{+}\right)^{-1}{\left[\phi_{2}^{+}\right]}\right]}\right]}\,.
        \end{align}
    \end{subequations}

\section{Arbitrary Chirping}\label{sec:appendix:Arbitrary_Chirping}
    \pati{Announcement}{ 
    }
    
    Here, we derive the interface trajectory, $z{\left[t\right]}$, corresponding to a prescribed chirping modulation profile, $\varphi{\left[t\right]}$ for the \emph{transmitted} wave. Although the analysis focuses on the transmitted component, the methodology is equally applicable to other scattered waves.

    \subsection{Interface Profile}\label{subsec:appendix:Interface_Profile}    
    \pati{Derivation Interface Profile}{ 
    }
    
    To impose a prescribed chirping profile on the transmitted wave, we require that its phase argument [Eq.~\eqref{eq:appendix:Subluminal_Scattering_Solutions_Transmitted_Wave}] matches the chirping profile:
    \begin{equation}\label{eq:appendix_Chirping_Profile}
    \frac{z{\left[t^{\star}\right]}}{u_{1}} - t^{\star} = \varphi{\left[x\right]}\,,
    \end{equation}
    where $t^{\star} = \left(f_{2}^{+}\right)^{-1}{\left[x\right]}$ denotes the scattering time at $x = z/u_{2} - t$ [Eq.~\eqref{eq:appendix:Subluminal_Change_of_Variables_Transmission}]. Directly solving Eq.~\eqref{eq:appendix_Chirping_Profile} for $z{\left[t\right]}$ is not straightforward, as both $z{\left[t^{\star}\right]}$ and $t^{\star}$ depend on the unknown interface profile. However, Eq.~\eqref{eq:appendix_Chirping_Profile}, together with the change of variables relating $z{\left[t^{\star}\right]}$ and $t^{\star}$ to $x$ [Eq.~\eqref{eq:appendix:Subluminal_Desired_Traveling_Waveform}], forms a system of equations    
    \begin{equation}\label{eq:appendix:Arbitrary_Chirping_System_of_Equations}
        \begin{cases}
            \dfrac{z{\left[t^{\star}\right]}}{u_{1}} - t^{\star} &= \varphi{\left[x\right]} \,, \vspace{1mm} \\
            \dfrac{z{\left[t^{\star}\right]}}{u_{2}} - t^{\star} &= x\,,
        \end{cases}
    \end{equation}
    which can be solved for $z{\left[t^{\star}\right]}$ and $t^{\star}$:
    \begin{equation}\label{eq:appendix:Arbitrary_Chirping_Solution_System_of_Equations}
        \begin{cases}
            z{\left[t^{\star}\right]} &= \dfrac{u_{1}u_{2}}{u_{1}-u_{2}}\left(x-\varphi{\left[x\right]}\right)\,, \vspace{1mm} \\
            t^{\star} &= \dfrac{u_{2}x - u_{1}\varphi{\left[x\right]}}{u_{1}-u_{2}}\,.
        \end{cases}
    \end{equation}
    Note that Eq.~\eqref{eq:appendix:Arbitrary_Chirping_Solution_System_of_Equations} provides $z{\left[t^{\star}\right]}$, whereas we seek $z{\left[t\right]}$. To express the interface equation of motion in terms of the laboratory time, we substitute the second equation Eq.~\eqref{eq:appendix:Arbitrary_Chirping_Solution_System_of_Equations} into the first, giving
    \begin{equation}\label{eq:appendix:Arbitrary_Chirping_Bare_Result}
        z{\left[\frac{u_{2}x-u_{1}\varphi{\left[x\right]}}{u_{1}-u_{2}}\right]} = \frac{u_{1}u_{2}}{u_{1}-u_{2}}\left(x-\varphi{\left[x\right]}\right)\,.
    \end{equation}
    Next, we introduce the change of variables:
    \begin{equation}\label{eq:appendix:Arbitrary_Chirping_Change_of_Variables}
        \frac{u_{2}x-u_{1}\varphi{\left[x\right]}}{u_{1}-u_{2}}\mapsto t\,,
    \end{equation}
    which implicitly defines $x{\left[t\right]}$ for a given chirping profile $\varphi{\left[t\right]}$. Substituting Eq.~\eqref{eq:appendix:Arbitrary_Chirping_Change_of_Variables} into Eq.~\eqref{eq:appendix:Arbitrary_Chirping_Bare_Result} yields the final expression for the interface profile,
    \begin{equation}\label{eq:appendix:Arbitrary_Chirping_Result_Interface_Profile}
        z{\left[t\right]} = \frac{u_{1}u_{2}}{u_{1}-u_{2}}\left(x{\left[t\right]} - \varphi{\left[x{\left[t\right]}\right]}\right)\,,
    \end{equation}
    where $x{\left[t\right]}$ solves Eq.~\eqref{eq:appendix:Arbitrary_Chirping_Change_of_Variables} and may require numerical methods for complex chirping profiles $\varphi{\left[x\right]}$.

    \subsection{Limitation}\label{subsec:appendix:Limitation}
    \pati{Derivation Limitation Arbitrary Chirping}{ 
    }
    
    There is a physical constraint on the choice of the chirping profile, $\varphi{\left[t\right]}$, as scattered transmitted waves only exist when $\beta{\left[t\right]} < 1/n_{2}$ (Fig.~\ref{fig:Overview}). This constraint translates into a condition on $\varphi{\left[t\right]}$ upon differentiating Eq.~\eqref{eq:appendix:Arbitrary_Chirping_Result_Interface_Profile} with respect to time and normalizing the result with respect to the speed of light in vacuum, which yields
    \begin{equation}\label{eq:appendix:Arbitrary_Chirping_First_Expression_Velocity}
        \beta{\left[t\right]} = \frac{1}{c}\frac{\dd{z}}{\dd{t}} = \frac{1}{c}\frac{u_{1}u_{2}}{u_{1}-u_{2}}\left(1-\dot{\varphi}{\left[x{\left[t\right]}\right]}\right)\dot{x}{\left[t\right]}\,,
    \end{equation}
    where dots indicate time derivatives. A corresponding expression for $\dot{x}{\left[t\right]}$ may be found by differentiating Eq.~\eqref{eq:appendix:Arbitrary_Chirping_Change_of_Variables} with respect to $t$ and solving for $\dot{x}{\left[t\right]}$, viz.,
    \begin{equation}\label{eq:appendix:Arbitrary_Chirping_Expression_dotx}
        \dot{x}{\left[t\right]} = \frac{u_{1}-u_{2}}{u_{2}-u_{1}\dot{\varphi}{\left[x{\left[t\right]}\right]}}\,.
    \end{equation}
    Substituting Eq.~\eqref{eq:appendix:Arbitrary_Chirping_Expression_dotx} into Eq.~\eqref{eq:appendix:Arbitrary_Chirping_First_Expression_Velocity} yields
    \begin{equation}
        \beta{\left[t\right]} = \frac{1-\dot{\varphi}{\left[x{\left[t\right]}\right]}}{n_{1}-n_{2}\dot{\varphi}{\left[x{\left[t\right]}\right]}} \,.
    \end{equation}
    Imposing $\beta{\left[t\right]} < 1/n_{2}$, we find the following condition on the choice of the chirping profile:
    \begin{equation}
        \dot{\varphi}{\left[x{\left[t\right]}\right]} > \frac{n_{1}}{n_{2}}\,.
    \end{equation}

\bibliography{main}

\begin{thebibliography}{65}%
\makeatletter
\providecommand \@ifxundefined [1]{%
 \@ifx{#1\undefined}
}%
\providecommand \@ifnum [1]{%
 \ifnum #1\expandafter \@firstoftwo
 \else \expandafter \@secondoftwo
 \fi
}%
\providecommand \@ifx [1]{%
 \ifx #1\expandafter \@firstoftwo
 \else \expandafter \@secondoftwo
 \fi
}%
\providecommand \natexlab [1]{#1}%
\providecommand \enquote  [1]{``#1''}%
\providecommand \bibnamefont  [1]{#1}%
\providecommand \bibfnamefont [1]{#1}%
\providecommand \citenamefont [1]{#1}%
\providecommand \href@noop [0]{\@secondoftwo}%
\providecommand \href [0]{\begingroup \@sanitize@url \@href}%
\providecommand \@href[1]{\@@startlink{#1}\@@href}%
\providecommand \@@href[1]{\endgroup#1\@@endlink}%
\providecommand \@sanitize@url [0]{\catcode `\\12\catcode `\$12\catcode `\&12\catcode `\#12\catcode `\^12\catcode `\_12\catcode `\%12\relax}%
\providecommand \@@startlink[1]{}%
\providecommand \@@endlink[0]{}%
\providecommand \url  [0]{\begingroup\@sanitize@url \@url }%
\providecommand \@url [1]{\endgroup\@href {#1}{\urlprefix }}%
\providecommand \urlprefix  [0]{URL }%
\providecommand \Eprint [0]{\href }%
\providecommand \doibase [0]{https://doi.org/}%
\providecommand \selectlanguage [0]{\@gobble}%
\providecommand \bibinfo  [0]{\@secondoftwo}%
\providecommand \bibfield  [0]{\@secondoftwo}%
\providecommand \translation [1]{[#1]}%
\providecommand \BibitemOpen [0]{}%
\providecommand \bibitemStop [0]{}%
\providecommand \bibitemNoStop [0]{.\EOS\space}%
\providecommand \EOS [0]{\spacefactor3000\relax}%
\providecommand \BibitemShut  [1]{\csname bibitem#1\endcsname}%
\let\auto@bib@innerbib\@empty
\bibitem [{\citenamefont {Caloz}\ and\ \citenamefont {Deck-L{\'e}ger}(2019)}]{Caloz2019a_ST_Metamaterials_PUB}%
  \BibitemOpen
  \bibfield  {author} {\bibinfo {author} {\bibfnamefont {C.}~\bibnamefont {Caloz}}\ and\ \bibinfo {author} {\bibfnamefont {Z.-L.}\ \bibnamefont {Deck-L{\'e}ger}},\ }\bibfield  {title} {\bibinfo {title} {{S}pacetime metamaterials---{P}art {I}: {G}eneral concepts},\ }\href@noop {} {\bibfield  {journal} {\bibinfo  {journal} {IEEE Trans. Antennas Propag.}\ }\textbf {\bibinfo {volume} {68}},\ \bibinfo {pages} {1569} (\bibinfo {year} {2019})}\BibitemShut {NoStop}%
\bibitem [{\citenamefont {{}Caloz}\ and\ \citenamefont {Deck-L{\'e}ger}(2019)}]{Caloz2019b_ST_Metamaterials_PUB}%
  \BibitemOpen
  \bibfield  {author} {\bibinfo {author} {\bibfnamefont {C.}~\bibnamefont {{}Caloz}}\ and\ \bibinfo {author} {\bibfnamefont {Z.-L.}\ \bibnamefont {Deck-L{\'e}ger}},\ }\bibfield  {title} {\bibinfo {title} {{S}pacetime metamaterials---{P}art {I}{I}: {T}heory and applications},\ }\href@noop {} {\bibfield  {journal} {\bibinfo  {journal} {IEEE Trans. Antennas Propag.}\ }\textbf {\bibinfo {volume} {68}},\ \bibinfo {pages} {1583} (\bibinfo {year} {2019})}\BibitemShut {NoStop}%
\bibitem [{\citenamefont {Caloz}\ \emph {et~al.}(2022)\citenamefont {Caloz}, \citenamefont {Deck-L{\'e}ger}, \citenamefont {Bahrami}, \citenamefont {Vicente},\ and\ \citenamefont {Li}}]{Caloz2022_GSTEMs_PUB}%
  \BibitemOpen
  \bibfield  {author} {\bibinfo {author} {\bibfnamefont {C.}~\bibnamefont {Caloz}}, \bibinfo {author} {\bibfnamefont {Z.-L.}\ \bibnamefont {Deck-L{\'e}ger}}, \bibinfo {author} {\bibfnamefont {A.}~\bibnamefont {Bahrami}}, \bibinfo {author} {\bibfnamefont {O.~C.}\ \bibnamefont {Vicente}},\ and\ \bibinfo {author} {\bibfnamefont {Z.}~\bibnamefont {Li}},\ }\bibfield  {title} {\bibinfo {title} {{G}eneralized space-time engineered modulation ({G}{S}{T}{E}{M}) metamaterials: {A} global and extended perspective},\ }\href@noop {} {\bibfield  {journal} {\bibinfo  {journal} {IEEE Antennas Propag. Mag.}\ }\textbf {\bibinfo {volume} {65}},\ \bibinfo {pages} {60} (\bibinfo {year} {2022})}\BibitemShut {NoStop}%
\bibitem [{\citenamefont {Morgenthaler}(1958)}]{Morgenthaler1958_TEM_PUB}%
  \BibitemOpen
  \bibfield  {author} {\bibinfo {author} {\bibfnamefont {F.~R.}\ \bibnamefont {Morgenthaler}},\ }\bibfield  {title} {\bibinfo {title} {{V}elocity modulation of electromagnetic waves},\ }\href@noop {} {\bibfield  {journal} {\bibinfo  {journal} {IRE Trans. Microw. Theory Tech.}\ }\textbf {\bibinfo {volume} {6}},\ \bibinfo {pages} {167} (\bibinfo {year} {1958})}\BibitemShut {NoStop}%
\bibitem [{\citenamefont {Galiffi}\ \emph {et~al.}(2022)\citenamefont {Galiffi}, \citenamefont {Tirole}, \citenamefont {Yin}, \citenamefont {Li}, \citenamefont {Vezzoli}, \citenamefont {Huidobro}, \citenamefont {Silveirinha}, \citenamefont {Sapienza}, \citenamefont {Al{\`u}},\ and\ \citenamefont {Pendry}}]{Pendry2022_Review_TEM_PUB}%
  \BibitemOpen
  \bibfield  {author} {\bibinfo {author} {\bibfnamefont {E.}~\bibnamefont {Galiffi}}, \bibinfo {author} {\bibfnamefont {R.}~\bibnamefont {Tirole}}, \bibinfo {author} {\bibfnamefont {S.}~\bibnamefont {Yin}}, \bibinfo {author} {\bibfnamefont {H.}~\bibnamefont {Li}}, \bibinfo {author} {\bibfnamefont {S.}~\bibnamefont {Vezzoli}}, \bibinfo {author} {\bibfnamefont {P.~A.}\ \bibnamefont {Huidobro}}, \bibinfo {author} {\bibfnamefont {M.~G.}\ \bibnamefont {Silveirinha}}, \bibinfo {author} {\bibfnamefont {R.}~\bibnamefont {Sapienza}}, \bibinfo {author} {\bibfnamefont {A.}~\bibnamefont {Al{\`u}}},\ and\ \bibinfo {author} {\bibfnamefont {J.~B.}\ \bibnamefont {Pendry}},\ }\bibfield  {title} {\bibinfo {title} {{P}hotonics of time-varying media},\ }\href@noop {} {\bibfield  {journal} {\bibinfo  {journal} {Adv. Photonics}\ }\textbf {\bibinfo {volume} {4}},\ \bibinfo {pages} {014002} (\bibinfo {year} {2022})}\BibitemShut {NoStop}%
\bibitem [{\citenamefont {Mostafa}\ \emph {et~al.}(2024)\citenamefont {Mostafa}, \citenamefont {Mirmoosa}, \citenamefont {Sidorenko}, \citenamefont {Asadchy},\ and\ \citenamefont {Tretyakov}}]{Mostafa2024_PUB}%
  \BibitemOpen
  \bibfield  {author} {\bibinfo {author} {\bibfnamefont {M.~H.}\ \bibnamefont {Mostafa}}, \bibinfo {author} {\bibfnamefont {M.~S.}\ \bibnamefont {Mirmoosa}}, \bibinfo {author} {\bibfnamefont {M.~S.}\ \bibnamefont {Sidorenko}}, \bibinfo {author} {\bibfnamefont {V.~S.}\ \bibnamefont {Asadchy}},\ and\ \bibinfo {author} {\bibfnamefont {S.~A.}\ \bibnamefont {Tretyakov}},\ }\bibfield  {title} {\bibinfo {title} {{T}emporal interfaces in complex electromagnetic materials: an overview},\ }\href@noop {} {\bibfield  {journal} {\bibinfo  {journal} {Opt. Mater. Express.}\ }\textbf {\bibinfo {volume} {14}},\ \bibinfo {pages} {1103} (\bibinfo {year} {2024})}\BibitemShut {NoStop}%
\bibitem [{\citenamefont {Bacot}\ \emph {et~al.}(2016)\citenamefont {Bacot}, \citenamefont {Labousse}, \citenamefont {Eddi}, \citenamefont {Fink},\ and\ \citenamefont {Fort}}]{Fink2016_TR_TEM_PUB}%
  \BibitemOpen
  \bibfield  {author} {\bibinfo {author} {\bibfnamefont {V.}~\bibnamefont {Bacot}}, \bibinfo {author} {\bibfnamefont {M.}~\bibnamefont {Labousse}}, \bibinfo {author} {\bibfnamefont {A.}~\bibnamefont {Eddi}}, \bibinfo {author} {\bibfnamefont {M.}~\bibnamefont {Fink}},\ and\ \bibinfo {author} {\bibfnamefont {E.}~\bibnamefont {Fort}},\ }\bibfield  {title} {\bibinfo {title} {{T}ime reversal and holography with spacetime transformations},\ }\href@noop {} {\bibfield  {journal} {\bibinfo  {journal} {Nat. Phys.}\ }\textbf {\bibinfo {volume} {12}},\ \bibinfo {pages} {972} (\bibinfo {year} {2016})}\BibitemShut {NoStop}%
\bibitem [{\citenamefont {Hayran}\ and\ \citenamefont {Monticone}(2024)}]{Monticone2024_Rozanov_Bound_TEM_PUB}%
  \BibitemOpen
  \bibfield  {author} {\bibinfo {author} {\bibfnamefont {Z.}~\bibnamefont {Hayran}}\ and\ \bibinfo {author} {\bibfnamefont {F.}~\bibnamefont {Monticone}},\ }\bibfield  {title} {\bibinfo {title} {{B}eyond the {R}ozanov bound on electromagnetic absorption via periodic temporal modulations},\ }\href@noop {} {\bibfield  {journal} {\bibinfo  {journal} {Phys. Rev. Appl.}\ }\textbf {\bibinfo {volume} {21}},\ \bibinfo {pages} {044007} (\bibinfo {year} {2024})}\BibitemShut {NoStop}%
\bibitem [{\citenamefont {Li}\ \emph {et~al.}(2019)\citenamefont {Li}, \citenamefont {Mekawy},\ and\ \citenamefont {Al{\`u}}}]{Alu2019_Chu_Limit_TEM_PUB}%
  \BibitemOpen
  \bibfield  {author} {\bibinfo {author} {\bibfnamefont {H.}~\bibnamefont {Li}}, \bibinfo {author} {\bibfnamefont {A.}~\bibnamefont {Mekawy}},\ and\ \bibinfo {author} {\bibfnamefont {A.}~\bibnamefont {Al{\`u}}},\ }\bibfield  {title} {\bibinfo {title} {{B}eyond {C}hu{'}s limit with {F}loquet impedance matching},\ }\href@noop {} {\bibfield  {journal} {\bibinfo  {journal} {Phys. Rev. Lett.}\ }\textbf {\bibinfo {volume} {123}},\ \bibinfo {pages} {164102} (\bibinfo {year} {2019})}\BibitemShut {NoStop}%
\bibitem [{\citenamefont {Shlivinski}\ and\ \citenamefont {Hadad}(2018)}]{Hadad2018_Bode-Fano_Bound_TEM_PUB}%
  \BibitemOpen
  \bibfield  {author} {\bibinfo {author} {\bibfnamefont {A.}~\bibnamefont {Shlivinski}}\ and\ \bibinfo {author} {\bibfnamefont {Y.}~\bibnamefont {Hadad}},\ }\bibfield  {title} {\bibinfo {title} {{B}eyond the {B}ode-{F}ano bound: {W}ideband impedance matching for short pulses using temporal switching of transmission-line parameters},\ }\href@noop {} {\bibfield  {journal} {\bibinfo  {journal} {Phys. Rev. Lett.}\ }\textbf {\bibinfo {volume} {121}},\ \bibinfo {pages} {204301} (\bibinfo {year} {2018})}\BibitemShut {NoStop}%
\bibitem [{\citenamefont {Pendry}(2024)}]{Pendry2024_Air_Cond_Phot_USTEM_PUB}%
  \BibitemOpen
  \bibfield  {author} {\bibinfo {author} {\bibfnamefont {J.~B.}\ \bibnamefont {Pendry}},\ }\bibfield  {title} {\bibinfo {title} {{A}ir conditioning for photons},\ }\href@noop {} {\bibfield  {journal} {\bibinfo  {journal} {Opt. Mater. Express.}\ }\textbf {\bibinfo {volume} {14}},\ \bibinfo {pages} {407} (\bibinfo {year} {2024})}\BibitemShut {NoStop}%
\bibitem [{\citenamefont {Akbarzadeh}\ \emph {et~al.}(2018)\citenamefont {Akbarzadeh}, \citenamefont {Chamanara},\ and\ \citenamefont {Caloz}}]{Caloz2018_Inverse_Prism_PUB}%
  \BibitemOpen
  \bibfield  {author} {\bibinfo {author} {\bibfnamefont {A.}~\bibnamefont {Akbarzadeh}}, \bibinfo {author} {\bibfnamefont {N.}~\bibnamefont {Chamanara}},\ and\ \bibinfo {author} {\bibfnamefont {C.}~\bibnamefont {Caloz}},\ }\bibfield  {title} {\bibinfo {title} {{I}nverse prism based on temporal discontinuity and spatial dispersion},\ }\href@noop {} {\bibfield  {journal} {\bibinfo  {journal} {Opt. Express.}\ }\textbf {\bibinfo {volume} {43}},\ \bibinfo {pages} {3297} (\bibinfo {year} {2018})}\BibitemShut {NoStop}%
\bibitem [{\citenamefont {Pacheco-Pe{\~n}a}\ and\ \citenamefont {Engheta}(2020{\natexlab{a}})}]{Engheta2020_Aiming_PUB}%
  \BibitemOpen
  \bibfield  {author} {\bibinfo {author} {\bibfnamefont {V.}~\bibnamefont {Pacheco-Pe{\~n}a}}\ and\ \bibinfo {author} {\bibfnamefont {N.}~\bibnamefont {Engheta}},\ }\bibfield  {title} {\bibinfo {title} {{T}emporal aiming},\ }\href@noop {} {\bibfield  {journal} {\bibinfo  {journal} {Light Sci. Appl.}\ }\textbf {\bibinfo {volume} {9}},\ \bibinfo {pages} {129} (\bibinfo {year} {2020}{\natexlab{a}})}\BibitemShut {NoStop}%
\bibitem [{\citenamefont {Pacheco-Pe{\~n}a}\ and\ \citenamefont {Engheta}(2020{\natexlab{b}})}]{Engheta2020_Coating_PUB}%
  \BibitemOpen
  \bibfield  {author} {\bibinfo {author} {\bibfnamefont {V.}~\bibnamefont {Pacheco-Pe{\~n}a}}\ and\ \bibinfo {author} {\bibfnamefont {N.}~\bibnamefont {Engheta}},\ }\bibfield  {title} {\bibinfo {title} {{A}ntireflection temporal coatings},\ }\href@noop {} {\bibfield  {journal} {\bibinfo  {journal} {Opt. Mater. Express.}\ }\textbf {\bibinfo {volume} {7}},\ \bibinfo {pages} {323} (\bibinfo {year} {2020}{\natexlab{b}})}\BibitemShut {NoStop}%
\bibitem [{\citenamefont {Tien}(1958)}]{Tien1958_PUB}%
  \BibitemOpen
  \bibfield  {author} {\bibinfo {author} {\bibfnamefont {P.~K.}\ \bibnamefont {Tien}},\ }\bibfield  {title} {\bibinfo {title} {{P}arametric amplification and frequency mixing in propagating circuits},\ }\href@noop {} {\bibfield  {journal} {\bibinfo  {journal} {J. Appl. Phys.}\ }\textbf {\bibinfo {volume} {29}},\ \bibinfo {pages} {1347} (\bibinfo {year} {1958})}\BibitemShut {NoStop}%
\bibitem [{\citenamefont {Mendon{\c{c}}a}\ \emph {et~al.}(2003)\citenamefont {Mendon{\c{c}}a}, \citenamefont {Martins},\ and\ \citenamefont {Guerreiro}}]{Guerreiro2003_PUB}%
  \BibitemOpen
  \bibfield  {author} {\bibinfo {author} {\bibfnamefont {J.~T.}\ \bibnamefont {Mendon{\c{c}}a}}, \bibinfo {author} {\bibfnamefont {A.~M.}\ \bibnamefont {Martins}},\ and\ \bibinfo {author} {\bibfnamefont {A.}~\bibnamefont {Guerreiro}},\ }\bibfield  {title} {\bibinfo {title} {{T}emporal beam splitter and temporal interference},\ }\href@noop {} {\bibfield  {journal} {\bibinfo  {journal} {Phys. Rev. A}\ }\textbf {\bibinfo {volume} {68}},\ \bibinfo {pages} {043801} (\bibinfo {year} {2003})}\BibitemShut {NoStop}%
\bibitem [{\citenamefont {Mendon{\c{c}}a}\ \emph {et~al.}(2000)\citenamefont {Mendon{\c{c}}a}, \citenamefont {Guerreiro},\ and\ \citenamefont {Martins}}]{Guerreiro2000_PUB}%
  \BibitemOpen
  \bibfield  {author} {\bibinfo {author} {\bibfnamefont {J.~T.}\ \bibnamefont {Mendon{\c{c}}a}}, \bibinfo {author} {\bibfnamefont {A.}~\bibnamefont {Guerreiro}},\ and\ \bibinfo {author} {\bibfnamefont {A.~M.}\ \bibnamefont {Martins}},\ }\bibfield  {title} {\bibinfo {title} {{Q}uantum theory of time refraction},\ }\href@noop {} {\bibfield  {journal} {\bibinfo  {journal} {Phys. Rev. A}\ }\textbf {\bibinfo {volume} {62}},\ \bibinfo {pages} {033805} (\bibinfo {year} {2000})}\BibitemShut {NoStop}%
\bibitem [{\citenamefont {Pacheco-Pe{\~n}a}\ and\ \citenamefont {Engheta}(2021)}]{Engheta2021_Brewster_PUB}%
  \BibitemOpen
  \bibfield  {author} {\bibinfo {author} {\bibfnamefont {V.}~\bibnamefont {Pacheco-Pe{\~n}a}}\ and\ \bibinfo {author} {\bibfnamefont {N.}~\bibnamefont {Engheta}},\ }\bibfield  {title} {\bibinfo {title} {{T}emporal equivalent of the {B}rewster angle},\ }\href@noop {} {\bibfield  {journal} {\bibinfo  {journal} {Phys. Rev. B}\ }\textbf {\bibinfo {volume} {104}},\ \bibinfo {pages} {214308} (\bibinfo {year} {2021})}\BibitemShut {NoStop}%
\bibitem [{\citenamefont {Mirmoosa}\ \emph {et~al.}(2024)\citenamefont {Mirmoosa}, \citenamefont {Mostafa}, \citenamefont {Norrman},\ and\ \citenamefont {Tretyakov}}]{Tretyakov2024_PUB}%
  \BibitemOpen
  \bibfield  {author} {\bibinfo {author} {\bibfnamefont {M.~S.}\ \bibnamefont {Mirmoosa}}, \bibinfo {author} {\bibfnamefont {M.~H.}\ \bibnamefont {Mostafa}}, \bibinfo {author} {\bibfnamefont {A.}~\bibnamefont {Norrman}},\ and\ \bibinfo {author} {\bibfnamefont {S.~A.}\ \bibnamefont {Tretyakov}},\ }\bibfield  {title} {\bibinfo {title} {{T}ime interfaces in bianisotropic media},\ }\href@noop {} {\bibfield  {journal} {\bibinfo  {journal} {Phys. Rev. Res.}\ }\textbf {\bibinfo {volume} {6}},\ \bibinfo {pages} {013334} (\bibinfo {year} {2024})}\BibitemShut {NoStop}%
\bibitem [{\citenamefont {Xu}\ \emph {et~al.}(2021)\citenamefont {Xu}, \citenamefont {Mai},\ and\ \citenamefont {Werner}}]{Werner2021_PUB}%
  \BibitemOpen
  \bibfield  {author} {\bibinfo {author} {\bibfnamefont {J.}~\bibnamefont {Xu}}, \bibinfo {author} {\bibfnamefont {W.}~\bibnamefont {Mai}},\ and\ \bibinfo {author} {\bibfnamefont {H.}~\bibnamefont {Werner}, \bibfnamefont {Douglas}},\ }\bibfield  {title} {\bibinfo {title} {{C}omplete polarization conversion using anisotropic temporal slabs},\ }\href@noop {} {\bibfield  {journal} {\bibinfo  {journal} {Opt. Express.}\ }\textbf {\bibinfo {volume} {46}},\ \bibinfo {pages} {1373} (\bibinfo {year} {2021})}\BibitemShut {NoStop}%
\bibitem [{\citenamefont {He}\ \emph {et~al.}(2023)\citenamefont {He}, \citenamefont {Shang}, \citenamefont {Qi}, \citenamefont {Bo},\ and\ \citenamefont {Li}}]{Huanan2023_PUB}%
  \BibitemOpen
  \bibfield  {author} {\bibinfo {author} {\bibfnamefont {H.}~\bibnamefont {He}}, \bibinfo {author} {\bibfnamefont {S.}~\bibnamefont {Shang}}, \bibinfo {author} {\bibfnamefont {J.}~\bibnamefont {Qi}}, \bibinfo {author} {\bibfnamefont {F.}~\bibnamefont {Bo}},\ and\ \bibinfo {author} {\bibfnamefont {H.}~\bibnamefont {Li}},\ }\bibfield  {title} {\bibinfo {title} {{F}araday rotation in nonreciprocal photonic time-crystals},\ }\href@noop {} {\bibfield  {journal} {\bibinfo  {journal} {Appl. Phys. Lett.}\ }\textbf {\bibinfo {volume} {122}},\ \bibinfo {pages} {051703} (\bibinfo {year} {2023})}\BibitemShut {NoStop}%
\bibitem [{\citenamefont {Li}\ \emph {et~al.}(2022)\citenamefont {Li}, \citenamefont {Yin},\ and\ \citenamefont {Al{\`{u}}}}]{Alu2022_PUB}%
  \BibitemOpen
  \bibfield  {author} {\bibinfo {author} {\bibfnamefont {H.}~\bibnamefont {Li}}, \bibinfo {author} {\bibfnamefont {S.}~\bibnamefont {Yin}},\ and\ \bibinfo {author} {\bibfnamefont {A.}~\bibnamefont {Al{\`{u}}}},\ }\bibfield  {title} {\bibinfo {title} {{N}onreciprocity and {F}araday rotation at time interfaces},\ }\href@noop {} {\bibfield  {journal} {\bibinfo  {journal} {Phys. Rev. Lett.}\ }\textbf {\bibinfo {volume} {128}},\ \bibinfo {pages} {173901} (\bibinfo {year} {2022})}\BibitemShut {NoStop}%
\bibitem [{\citenamefont {Cassedy}\ and\ \citenamefont {Oliner}(1963)}]{Cassedy1963_PUB}%
  \BibitemOpen
  \bibfield  {author} {\bibinfo {author} {\bibfnamefont {E.~S.}\ \bibnamefont {Cassedy}}\ and\ \bibinfo {author} {\bibfnamefont {A.~A.}\ \bibnamefont {Oliner}},\ }\bibfield  {title} {\bibinfo {title} {{D}ispersion relations in time-space periodic media: Part {I}---{S}table interactions},\ }\href@noop {} {\bibfield  {journal} {\bibinfo  {journal} {Proc. IEEE}\ }\textbf {\bibinfo {volume} {51}},\ \bibinfo {pages} {1342} (\bibinfo {year} {1963})}\BibitemShut {NoStop}%
\bibitem [{\citenamefont {Cassedy}(1967)}]{Cassedy1967_PUB}%
  \BibitemOpen
  \bibfield  {author} {\bibinfo {author} {\bibfnamefont {E.~S.}\ \bibnamefont {Cassedy}},\ }\bibfield  {title} {\bibinfo {title} {{D}ispersion relations in time-space periodic media: {P}art {I}{I}---{U}nstable interactions},\ }\href@noop {} {\bibfield  {journal} {\bibinfo  {journal} {Proc. IEEE}\ }\textbf {\bibinfo {volume} {55}},\ \bibinfo {pages} {1154} (\bibinfo {year} {1967})}\BibitemShut {NoStop}%
\bibitem [{\citenamefont {Bolotovski{\v{i}}}\ and\ \citenamefont {Ginzburg}(1972)}]{Bolotovskii1972_PUB}%
  \BibitemOpen
  \bibfield  {author} {\bibinfo {author} {\bibfnamefont {B.~M.}\ \bibnamefont {Bolotovski{\v{i}}}}\ and\ \bibinfo {author} {\bibfnamefont {V.~L.}\ \bibnamefont {Ginzburg}},\ }\bibfield  {title} {\bibinfo {title} {{T}he {V}avilov-{C}erenkov effect and the {D}oppler effect in the motion of sources with superluminal velocity in vacuum},\ }\href@noop {} {\bibfield  {journal} {\bibinfo  {journal} {Sov. Phys. Uspekhi}\ }\textbf {\bibinfo {volume} {15}},\ \bibinfo {pages} {184} (\bibinfo {year} {1972})}\BibitemShut {NoStop}%
\bibitem [{\citenamefont {Deck-L{\'e}ger}\ \emph {et~al.}(2019)\citenamefont {Deck-L{\'e}ger}, \citenamefont {Chamanara}, \citenamefont {Skorobogatiy}, \citenamefont {Silveirinha},\ and\ \citenamefont {Caloz}}]{Deck-Leger2019_Uni_Vel_PUB}%
  \BibitemOpen
  \bibfield  {author} {\bibinfo {author} {\bibfnamefont {Z.-L.}\ \bibnamefont {Deck-L{\'e}ger}}, \bibinfo {author} {\bibfnamefont {N.}~\bibnamefont {Chamanara}}, \bibinfo {author} {\bibfnamefont {M.}~\bibnamefont {Skorobogatiy}}, \bibinfo {author} {\bibfnamefont {M.~G.}\ \bibnamefont {Silveirinha}},\ and\ \bibinfo {author} {\bibfnamefont {C.}~\bibnamefont {Caloz}},\ }\bibfield  {title} {\bibinfo {title} {{U}niform-velocity spacetime crystals},\ }\href@noop {} {\bibfield  {journal} {\bibinfo  {journal} {Adv. Photonics}\ }\textbf {\bibinfo {volume} {1}},\ \bibinfo {pages} {56002} (\bibinfo {year} {2019})}\BibitemShut {NoStop}%
\bibitem [{\citenamefont {Gaafar}\ \emph {et~al.}(2019)\citenamefont {Gaafar}, \citenamefont {Baba}, \citenamefont {Eich},\ and\ \citenamefont {Petrov}}]{Gaafar2019_PUB}%
  \BibitemOpen
  \bibfield  {author} {\bibinfo {author} {\bibfnamefont {M.~A.}\ \bibnamefont {Gaafar}}, \bibinfo {author} {\bibfnamefont {T.}~\bibnamefont {Baba}}, \bibinfo {author} {\bibfnamefont {M.}~\bibnamefont {Eich}},\ and\ \bibinfo {author} {\bibfnamefont {A.}~\bibnamefont {Petrov}},\ }\bibfield  {title} {\bibinfo {title} {{F}ront-induced transitions},\ }\href@noop {} {\bibfield  {journal} {\bibinfo  {journal} {Nat. Photonics.}\ }\textbf {\bibinfo {volume} {13}},\ \bibinfo {pages} {737} (\bibinfo {year} {2019})}\BibitemShut {NoStop}%
\bibitem [{\citenamefont {Granatstein}\ \emph {et~al.}(1976)\citenamefont {Granatstein}, \citenamefont {Sprangle}, \citenamefont {Parker}, \citenamefont {Pasour}, \citenamefont {Herndon}, \citenamefont {Schlesinger},\ and\ \citenamefont {Seftor}}]{Seftor1976_PUB}%
  \BibitemOpen
  \bibfield  {author} {\bibinfo {author} {\bibfnamefont {V.~L.}\ \bibnamefont {Granatstein}}, \bibinfo {author} {\bibfnamefont {P.}~\bibnamefont {Sprangle}}, \bibinfo {author} {\bibfnamefont {R.~K.}\ \bibnamefont {Parker}}, \bibinfo {author} {\bibfnamefont {J.}~\bibnamefont {Pasour}}, \bibinfo {author} {\bibfnamefont {M.}~\bibnamefont {Herndon}}, \bibinfo {author} {\bibfnamefont {S.~P.}\ \bibnamefont {Schlesinger}},\ and\ \bibinfo {author} {\bibfnamefont {J.~L.}\ \bibnamefont {Seftor}},\ }\bibfield  {title} {\bibinfo {title} {{R}ealization of a relativistic mirror: {E}lectromagnetic backscattering from the front of a magnetized relativistic electron beam},\ }\href@noop {} {\bibfield  {journal} {\bibinfo  {journal} {Phys. Rev. A}\ }\textbf {\bibinfo {volume} {14}},\ \bibinfo {pages} {1194} (\bibinfo {year} {1976})}\BibitemShut {NoStop}%
\bibitem [{\citenamefont {Lampe}\ \emph {et~al.}(1978)\citenamefont {Lampe}, \citenamefont {Ott},\ and\ \citenamefont {Walker}}]{Lampe1978_PUB}%
  \BibitemOpen
  \bibfield  {author} {\bibinfo {author} {\bibfnamefont {M.}~\bibnamefont {Lampe}}, \bibinfo {author} {\bibfnamefont {E.}~\bibnamefont {Ott}},\ and\ \bibinfo {author} {\bibfnamefont {J.~H.}\ \bibnamefont {Walker}},\ }\bibfield  {title} {\bibinfo {title} {{I}nteraction of electromagnetic waves with a moving ionization front},\ }\href@noop {} {\bibfield  {journal} {\bibinfo  {journal} {Phys. Fluids}\ }\textbf {\bibinfo {volume} {21}},\ \bibinfo {pages} {42} (\bibinfo {year} {1978})}\BibitemShut {NoStop}%
\bibitem [{\citenamefont {Hadad}\ and\ \citenamefont {Sounas}(2024)}]{Sounas2024_PUB}%
  \BibitemOpen
  \bibfield  {author} {\bibinfo {author} {\bibfnamefont {Y.}~\bibnamefont {Hadad}}\ and\ \bibinfo {author} {\bibfnamefont {D.}~\bibnamefont {Sounas}},\ }\bibfield  {title} {\bibinfo {title} {{S}pace-time modulated loaded-wire metagratings for magnetless nonreciprocity and near-complete frequency conversion},\ }\href@noop {} {\bibfield  {journal} {\bibinfo  {journal} {Opt. Mater. Express.}\ }\textbf {\bibinfo {volume} {14}},\ \bibinfo {pages} {1295} (\bibinfo {year} {2024})}\BibitemShut {NoStop}%
\bibitem [{\citenamefont {Taravati}\ \emph {et~al.}(2017)\citenamefont {Taravati}, \citenamefont {Chamanara},\ and\ \citenamefont {Caloz}}]{Caloz2017_Nonreciprocity_PUB}%
  \BibitemOpen
  \bibfield  {author} {\bibinfo {author} {\bibfnamefont {S.}~\bibnamefont {Taravati}}, \bibinfo {author} {\bibfnamefont {N.}~\bibnamefont {Chamanara}},\ and\ \bibinfo {author} {\bibfnamefont {C.}~\bibnamefont {Caloz}},\ }\bibfield  {title} {\bibinfo {title} {{N}onreciprocal electromagnetic scattering from a periodically space-time modulated slab and application to a quasisonic isolator},\ }\href@noop {} {\bibfield  {journal} {\bibinfo  {journal} {Phys. Rev. B}\ }\textbf {\bibinfo {volume} {96}},\ \bibinfo {pages} {165144} (\bibinfo {year} {2017})}\BibitemShut {NoStop}%
\bibitem [{\citenamefont {Estep}\ \emph {et~al.}(2014)\citenamefont {Estep}, \citenamefont {Sounas}, \citenamefont {Soric},\ and\ \citenamefont {Al{\`{u}}}}]{Alu2014_PUB}%
  \BibitemOpen
  \bibfield  {author} {\bibinfo {author} {\bibfnamefont {N.~A.}\ \bibnamefont {Estep}}, \bibinfo {author} {\bibfnamefont {D.~L.}\ \bibnamefont {Sounas}}, \bibinfo {author} {\bibfnamefont {J.}~\bibnamefont {Soric}},\ and\ \bibinfo {author} {\bibfnamefont {A.}~\bibnamefont {Al{\`{u}}}},\ }\bibfield  {title} {\bibinfo {title} {{M}agnetic-free non-reciprocity and isolation based on parametrically modulated coupled-resonator loops},\ }\href@noop {} {\bibfield  {journal} {\bibinfo  {journal} {Nat. Phys.}\ }\textbf {\bibinfo {volume} {10}},\ \bibinfo {pages} {923} (\bibinfo {year} {2014})}\BibitemShut {NoStop}%
\bibitem [{\citenamefont {Deck-L{\'e}ger}\ \emph {et~al.}(2018)\citenamefont {Deck-L{\'e}ger}, \citenamefont {Akbarzadeh},\ and\ \citenamefont {Caloz}}]{Deck-Leger2018_PUB}%
  \BibitemOpen
  \bibfield  {author} {\bibinfo {author} {\bibfnamefont {Z.-L.}\ \bibnamefont {Deck-L{\'e}ger}}, \bibinfo {author} {\bibfnamefont {A.}~\bibnamefont {Akbarzadeh}},\ and\ \bibinfo {author} {\bibfnamefont {C.}~\bibnamefont {Caloz}},\ }\bibfield  {title} {\bibinfo {title} {{W}ave deflection and shifted refocusing in a medium modulated by a superluminal rectangular pulse},\ }\href@noop {} {\bibfield  {journal} {\bibinfo  {journal} {Phys. Rev. B}\ }\textbf {\bibinfo {volume} {97}},\ \bibinfo {pages} {104305} (\bibinfo {year} {2018})}\BibitemShut {NoStop}%
\bibitem [{\citenamefont {Taravati}\ and\ \citenamefont {Eleftheriades}(2019)}]{Eleftheriades2019_PUB}%
  \BibitemOpen
  \bibfield  {author} {\bibinfo {author} {\bibfnamefont {S.}~\bibnamefont {Taravati}}\ and\ \bibinfo {author} {\bibfnamefont {G.~V.}\ \bibnamefont {Eleftheriades}},\ }\bibfield  {title} {\bibinfo {title} {{G}eneralized space-time-periodic diffraction gratings: {T}heory and applications},\ }\href@noop {} {\bibfield  {journal} {\bibinfo  {journal} {Phys. Rev. Appl.}\ }\textbf {\bibinfo {volume} {12}},\ \bibinfo {pages} {024026} (\bibinfo {year} {2019})}\BibitemShut {NoStop}%
\bibitem [{\citenamefont {Chamanara}\ \emph {et~al.}(2017)\citenamefont {Chamanara}, \citenamefont {Taravati}, \citenamefont {Deck-L{\'e}ger},\ and\ \citenamefont {Caloz}}]{Caloz2017_Isolation_PUB}%
  \BibitemOpen
  \bibfield  {author} {\bibinfo {author} {\bibfnamefont {N.}~\bibnamefont {Chamanara}}, \bibinfo {author} {\bibfnamefont {S.}~\bibnamefont {Taravati}}, \bibinfo {author} {\bibfnamefont {Z.-L.}\ \bibnamefont {Deck-L{\'e}ger}},\ and\ \bibinfo {author} {\bibfnamefont {C.}~\bibnamefont {Caloz}},\ }\bibfield  {title} {\bibinfo {title} {{O}ptical isolation based on space-time engineered asymmetric photonic band gaps},\ }\href@noop {} {\bibfield  {journal} {\bibinfo  {journal} {Phys. Rev. B}\ }\textbf {\bibinfo {volume} {96}},\ \bibinfo {pages} {155409} (\bibinfo {year} {2017})}\BibitemShut {NoStop}%
\bibitem [{\citenamefont {Huidobro}\ \emph {et~al.}(2019)\citenamefont {Huidobro}, \citenamefont {Galiffi}, \citenamefont {Guenneau}, \citenamefont {Craster},\ and\ \citenamefont {Pendry}}]{Pendry2019_PUB}%
  \BibitemOpen
  \bibfield  {author} {\bibinfo {author} {\bibfnamefont {P.~A.}\ \bibnamefont {Huidobro}}, \bibinfo {author} {\bibfnamefont {E.}~\bibnamefont {Galiffi}}, \bibinfo {author} {\bibfnamefont {S.}~\bibnamefont {Guenneau}}, \bibinfo {author} {\bibfnamefont {R.~V.}\ \bibnamefont {Craster}},\ and\ \bibinfo {author} {\bibfnamefont {J.~B.}\ \bibnamefont {Pendry}},\ }\bibfield  {title} {\bibinfo {title} {{F}resnel drag in space-time-modulated metamaterials},\ }\href@noop {} {\bibfield  {journal} {\bibinfo  {journal} {Proc. Natl. Acad. Sci. USA}\ }\textbf {\bibinfo {volume} {116}},\ \bibinfo {pages} {24943} (\bibinfo {year} {2019})}\BibitemShut {NoStop}%
\bibitem [{\citenamefont {Huidobro}\ \emph {et~al.}(2021)\citenamefont {Huidobro}, \citenamefont {Silveirinha}, \citenamefont {Galiffi},\ and\ \citenamefont {Pendry}}]{Pendry2021_Homogenization_PUB}%
  \BibitemOpen
  \bibfield  {author} {\bibinfo {author} {\bibfnamefont {P.~A.}\ \bibnamefont {Huidobro}}, \bibinfo {author} {\bibfnamefont {M.~G.}\ \bibnamefont {Silveirinha}}, \bibinfo {author} {\bibfnamefont {E.}~\bibnamefont {Galiffi}},\ and\ \bibinfo {author} {\bibfnamefont {J.~B.}\ \bibnamefont {Pendry}},\ }\bibfield  {title} {\bibinfo {title} {{H}omogenization theory of space-time metamaterials},\ }\href@noop {} {\bibfield  {journal} {\bibinfo  {journal} {Phys. Rev. Appl.}\ }\textbf {\bibinfo {volume} {16}},\ \bibinfo {pages} {014044} (\bibinfo {year} {2021})}\BibitemShut {NoStop}%
\bibitem [{\citenamefont {Reed}\ \emph {et~al.}(2003)\citenamefont {Reed}, \citenamefont {Solja{\v{c}}i{\'c}},\ and\ \citenamefont {Joannopoulos}}]{Joannopoulos2003_PUB}%
  \BibitemOpen
  \bibfield  {author} {\bibinfo {author} {\bibfnamefont {E.~J.}\ \bibnamefont {Reed}}, \bibinfo {author} {\bibfnamefont {M.}~\bibnamefont {Solja{\v{c}}i{\'c}}},\ and\ \bibinfo {author} {\bibfnamefont {J.~D.}\ \bibnamefont {Joannopoulos}},\ }\bibfield  {title} {\bibinfo {title} {{C}olor of shock waves in photonic crystals},\ }\href@noop {} {\bibfield  {journal} {\bibinfo  {journal} {Phys. Rev. Lett.}\ }\textbf {\bibinfo {volume} {90}},\ \bibinfo {pages} {203904} (\bibinfo {year} {2003})}\BibitemShut {NoStop}%
\bibitem [{\citenamefont {Philbin}\ \emph {et~al.}(2008)\citenamefont {Philbin}, \citenamefont {Kuklewicz}, \citenamefont {Robertson}, \citenamefont {Hill}, \citenamefont {K{\"{o}}nig},\ and\ \citenamefont {Leonhardt}}]{Leonhardt2008_PUB}%
  \BibitemOpen
  \bibfield  {author} {\bibinfo {author} {\bibfnamefont {T.~G.}\ \bibnamefont {Philbin}}, \bibinfo {author} {\bibfnamefont {C.}~\bibnamefont {Kuklewicz}}, \bibinfo {author} {\bibfnamefont {S.}~\bibnamefont {Robertson}}, \bibinfo {author} {\bibfnamefont {S.}~\bibnamefont {Hill}}, \bibinfo {author} {\bibfnamefont {F.}~\bibnamefont {K{\"{o}}nig}},\ and\ \bibinfo {author} {\bibfnamefont {U.}~\bibnamefont {Leonhardt}},\ }\bibfield  {title} {\bibinfo {title} {{F}iber-optical analog of the event horizon},\ }\href@noop {} {\bibfield  {journal} {\bibinfo  {journal} {Science}\ }\textbf {\bibinfo {volume} {319}},\ \bibinfo {pages} {1367} (\bibinfo {year} {2008})}\BibitemShut {NoStop}%
\bibitem [{\citenamefont {Drori}\ \emph {et~al.}(2019)\citenamefont {Drori}, \citenamefont {Rosenberg}, \citenamefont {Bermudez}, \citenamefont {Silberberg},\ and\ \citenamefont {Leonhardt}}]{Leonhardt2019_PUB}%
  \BibitemOpen
  \bibfield  {author} {\bibinfo {author} {\bibfnamefont {J.}~\bibnamefont {Drori}}, \bibinfo {author} {\bibfnamefont {Y.}~\bibnamefont {Rosenberg}}, \bibinfo {author} {\bibfnamefont {D.}~\bibnamefont {Bermudez}}, \bibinfo {author} {\bibfnamefont {Y.}~\bibnamefont {Silberberg}},\ and\ \bibinfo {author} {\bibfnamefont {U.}~\bibnamefont {Leonhardt}},\ }\bibfield  {title} {\bibinfo {title} {{O}bservation of stimulated {H}awking radiation in an optical analogue},\ }\href@noop {} {\bibfield  {journal} {\bibinfo  {journal} {Phys. Rev. Lett.}\ }\textbf {\bibinfo {volume} {122}},\ \bibinfo {pages} {010404} (\bibinfo {year} {2019})}\BibitemShut {NoStop}%
\bibitem [{\citenamefont {Li}\ \emph {et~al.}(2023)\citenamefont {Li}, \citenamefont {Ma}, \citenamefont {Bahrami}, \citenamefont {Deck-L{\'e}ger},\ and\ \citenamefont {Caloz}}]{Li2023_Fresnel_Prism_PUB}%
  \BibitemOpen
  \bibfield  {author} {\bibinfo {author} {\bibfnamefont {Z.}~\bibnamefont {Li}}, \bibinfo {author} {\bibfnamefont {X.}~\bibnamefont {Ma}}, \bibinfo {author} {\bibfnamefont {A.}~\bibnamefont {Bahrami}}, \bibinfo {author} {\bibfnamefont {Z.-L.}\ \bibnamefont {Deck-L{\'e}ger}},\ and\ \bibinfo {author} {\bibfnamefont {C.}~\bibnamefont {Caloz}},\ }\bibfield  {title} {\bibinfo {title} {{S}pace-time {F}resnel prism},\ }\href@noop {} {\bibfield  {journal} {\bibinfo  {journal} {Phys. Rev. Appl.}\ }\textbf {\bibinfo {volume} {20}},\ \bibinfo {pages} {054029} (\bibinfo {year} {2023})}\BibitemShut {NoStop}%
\bibitem [{\citenamefont {Pendry}\ \emph {et~al.}(2021)\citenamefont {Pendry}, \citenamefont {Galiffi},\ and\ \citenamefont {Huidobro}}]{Pendry2021_Gain_PUB}%
  \BibitemOpen
  \bibfield  {author} {\bibinfo {author} {\bibfnamefont {J.~B.}\ \bibnamefont {Pendry}}, \bibinfo {author} {\bibfnamefont {E.}~\bibnamefont {Galiffi}},\ and\ \bibinfo {author} {\bibfnamefont {P.~A.}\ \bibnamefont {Huidobro}},\ }\bibfield  {title} {\bibinfo {title} {{G}ain in time-dependent media--a new mechanism},\ }\href@noop {} {\bibfield  {journal} {\bibinfo  {journal} {J. Opt. Soc. Am. B}\ }\textbf {\bibinfo {volume} {38}},\ \bibinfo {pages} {3360} (\bibinfo {year} {2021})}\BibitemShut {NoStop}%
\bibitem [{\citenamefont {Yang}\ \emph {et~al.}(2023)\citenamefont {Yang}, \citenamefont {Hu}, \citenamefont {Li},\ and\ \citenamefont {Luo}}]{Luo2023_PUB}%
  \BibitemOpen
  \bibfield  {author} {\bibinfo {author} {\bibfnamefont {Q.}~\bibnamefont {Yang}}, \bibinfo {author} {\bibfnamefont {H.}~\bibnamefont {Hu}}, \bibinfo {author} {\bibfnamefont {X.}~\bibnamefont {Li}},\ and\ \bibinfo {author} {\bibfnamefont {Y.}~\bibnamefont {Luo}},\ }\bibfield  {title} {\bibinfo {title} {{C}ascaded parametric amplification based on spatiotemporal modulations},\ }\href@noop {} {\bibfield  {journal} {\bibinfo  {journal} {Photonics Res.}\ }\textbf {\bibinfo {volume} {11}},\ \bibinfo {pages} {B125} (\bibinfo {year} {2023})}\BibitemShut {NoStop}%
\bibitem [{\citenamefont {Bahrami}\ \emph {et~al.}(2025{\natexlab{a}})\citenamefont {Bahrami}, \citenamefont {De~Kinder}, \citenamefont {Li},\ and\ \citenamefont {Caloz}}]{Bahrami2025_Wedges_USTEM_PUB}%
  \BibitemOpen
  \bibfield  {author} {\bibinfo {author} {\bibfnamefont {A.}~\bibnamefont {Bahrami}}, \bibinfo {author} {\bibfnamefont {K.}~\bibnamefont {De~Kinder}}, \bibinfo {author} {\bibfnamefont {Z.}~\bibnamefont {Li}},\ and\ \bibinfo {author} {\bibfnamefont {C.}~\bibnamefont {Caloz}},\ }\bibfield  {title} {\bibinfo {title} {{S}pace-time wedges},\ }\href@noop {} {\bibfield  {journal} {\bibinfo  {journal} {Nanophotonics}\ } (\bibinfo {year} {2025}{\natexlab{a}})}\BibitemShut {NoStop}%
\bibitem [{\citenamefont {De~Kinder}\ \emph {et~al.}(2025)\citenamefont {De~Kinder}, \citenamefont {Bahrami},\ and\ \citenamefont {Caloz}}]{DeKinder2025_DoPA_USTEM}%
  \BibitemOpen
  \bibfield  {author} {\bibinfo {author} {\bibfnamefont {K.}~\bibnamefont {De~Kinder}}, \bibinfo {author} {\bibfnamefont {A.}~\bibnamefont {Bahrami}},\ and\ \bibinfo {author} {\bibfnamefont {C.}~\bibnamefont {Caloz}},\ }\href@noop {} {\bibinfo {title} {{D}oppler pulse amplification}} (\bibinfo {year} {2025}),\ \Eprint {https://arxiv.org/abs/2506.01447} {arXiv:2506.01447} \BibitemShut {NoStop}%
\bibitem [{\citenamefont {Ostrovski{\v{i}}}(1975)}]{Ostrovskii1975_Lens_PUB}%
  \BibitemOpen
  \bibfield  {author} {\bibinfo {author} {\bibfnamefont {L.~A.}\ \bibnamefont {Ostrovski{\v{i}}}},\ }\bibfield  {title} {\bibinfo {title} {{A} boundary with accelerating motion as a ``space-time lens''},\ }\href@noop {} {\bibfield  {journal} {\bibinfo  {journal} {Radiophys. Quantum Electron}\ }\textbf {\bibinfo {volume} {18}},\ \bibinfo {pages} {456} (\bibinfo {year} {1975})}\BibitemShut {NoStop}%
\bibitem [{\citenamefont {Bahrami}\ \emph {et~al.}(2025{\natexlab{b}})\citenamefont {Bahrami}, \citenamefont {De~Kinder},\ and\ \citenamefont {Caloz}}]{Bahrami2025_Pulse_Shap}%
  \BibitemOpen
  \bibfield  {author} {\bibinfo {author} {\bibfnamefont {A.}~\bibnamefont {Bahrami}}, \bibinfo {author} {\bibfnamefont {K.}~\bibnamefont {De~Kinder}},\ and\ \bibinfo {author} {\bibfnamefont {C.}~\bibnamefont {Caloz}},\ }\bibfield  {title} {\bibinfo {title} {{A}rbitrary pulse shaping using accelerated interfaces},\ }\href@noop {} {\bibfield  {journal} {\bibinfo  {journal} {Engineering Archive}\ } (\bibinfo {year} {2025}{\natexlab{b}})}\BibitemShut {NoStop}%
\bibitem [{\citenamefont {Sloan}\ \emph {et~al.}(2022)\citenamefont {Sloan}, \citenamefont {Rivera}, \citenamefont {Joannopoulos},\ and\ \citenamefont {Solja{\v{c}}i{\'c}}}]{Sloan2022_PUB}%
  \BibitemOpen
  \bibfield  {author} {\bibinfo {author} {\bibfnamefont {J.}~\bibnamefont {Sloan}}, \bibinfo {author} {\bibfnamefont {N.}~\bibnamefont {Rivera}}, \bibinfo {author} {\bibfnamefont {J.~D.}\ \bibnamefont {Joannopoulos}},\ and\ \bibinfo {author} {\bibfnamefont {M.}~\bibnamefont {Solja{\v{c}}i{\'c}}},\ }\bibfield  {title} {\bibinfo {title} {{C}ontrolling two-photon emission from superluminal and accelerating index perturbations},\ }\href@noop {} {\bibfield  {journal} {\bibinfo  {journal} {Nat. Phys.}\ }\textbf {\bibinfo {volume} {18}},\ \bibinfo {pages} {67} (\bibinfo {year} {2022})}\BibitemShut {NoStop}%
\bibitem [{\citenamefont {Bahrami}\ \emph {et~al.}(2024)\citenamefont {Bahrami}, \citenamefont {Deck-L{\'e}ger}, \citenamefont {Li},\ and\ \citenamefont {Caloz}}]{Bahrami2023_FDTD_PUB}%
  \BibitemOpen
  \bibfield  {author} {\bibinfo {author} {\bibfnamefont {A.}~\bibnamefont {Bahrami}}, \bibinfo {author} {\bibfnamefont {Z.-L.}\ \bibnamefont {Deck-L{\'e}ger}}, \bibinfo {author} {\bibfnamefont {Z.}~\bibnamefont {Li}},\ and\ \bibinfo {author} {\bibfnamefont {C.}~\bibnamefont {Caloz}},\ }\bibfield  {title} {\bibinfo {title} {{A} generalized {F}{D}{T}{D} scheme for moving electromagnetic structures with arbitrary space-time configurations},\ }\href@noop {} {\bibfield  {journal} {\bibinfo  {journal} {IEEE Trans. Antennas Propag.}\ }\textbf {\bibinfo {volume} {72}},\ \bibinfo {pages} {1721} (\bibinfo {year} {2024})}\BibitemShut {NoStop}%
\bibitem [{\citenamefont {Bahrami}\ \emph {et~al.}(2023)\citenamefont {Bahrami}, \citenamefont {Deck-L{\'e}ger},\ and\ \citenamefont {Caloz}}]{Bahrami2023_ASTEMs_PUB}%
  \BibitemOpen
  \bibfield  {author} {\bibinfo {author} {\bibfnamefont {A.}~\bibnamefont {Bahrami}}, \bibinfo {author} {\bibfnamefont {Z.-L.}\ \bibnamefont {Deck-L{\'e}ger}},\ and\ \bibinfo {author} {\bibfnamefont {C.}~\bibnamefont {Caloz}},\ }\bibfield  {title} {\bibinfo {title} {{E}lectrodynamics of accelerated-modulation space-time metamaterials},\ }\href@noop {} {\bibfield  {journal} {\bibinfo  {journal} {Phys. Rev. Appl.}\ }\textbf {\bibinfo {volume} {19}},\ \bibinfo {pages} {054044} (\bibinfo {year} {2023})}\BibitemShut {NoStop}%
\bibitem [{\citenamefont {Van~Bladel}(2012)}]{Bladel2012_BOOK}%
  \BibitemOpen
  \bibfield  {author} {\bibinfo {author} {\bibfnamefont {J.}~\bibnamefont {Van~Bladel}},\ }\href@noop {} {\emph {\bibinfo {title} {{R}elativity and {E}ngineering}}},\ Vol.~\bibinfo {volume} {15}\ (\bibinfo  {publisher} {Springer Science \& Business Media},\ \bibinfo {year} {2012})\BibitemShut {NoStop}%
\bibitem [{\citenamefont {Misner}\ \emph {et~al.}(1973)\citenamefont {Misner}, \citenamefont {Thorne},\ and\ \citenamefont {Wheeler}}]{Misner1973_GR_BOOK}%
  \BibitemOpen
  \bibfield  {author} {\bibinfo {author} {\bibfnamefont {C.~W.}\ \bibnamefont {Misner}}, \bibinfo {author} {\bibfnamefont {K.~S.}\ \bibnamefont {Thorne}},\ and\ \bibinfo {author} {\bibfnamefont {J.~A.}\ \bibnamefont {Wheeler}},\ }\href@noop {} {\emph {\bibinfo {title} {{G}ravitation}}}\ (\bibinfo  {publisher} {W. H. Freeman},\ \bibinfo {year} {1973})\BibitemShut {NoStop}%
\bibitem [{\citenamefont {Caloz}\ \emph {et~al.}(2013)\citenamefont {Caloz}, \citenamefont {Gupta}, \citenamefont {Zhang},\ and\ \citenamefont {Nikfal}}]{Caloz2013_PUB}%
  \BibitemOpen
  \bibfield  {author} {\bibinfo {author} {\bibfnamefont {C.}~\bibnamefont {Caloz}}, \bibinfo {author} {\bibfnamefont {S.}~\bibnamefont {Gupta}}, \bibinfo {author} {\bibfnamefont {Q.}~\bibnamefont {Zhang}},\ and\ \bibinfo {author} {\bibfnamefont {B.}~\bibnamefont {Nikfal}},\ }\bibfield  {title} {\bibinfo {title} {{A}nalog signal processing},\ }\href@noop {} {\bibfield  {journal} {\bibinfo  {journal} {IEEE Microw. Mag.}\ }\textbf {\bibinfo {volume} {14}},\ \bibinfo {pages} {87} (\bibinfo {year} {2013})}\BibitemShut {NoStop}%
\bibitem [{\citenamefont {Strickland}\ and\ \citenamefont {Mourou}(1985)}]{Strickland1985_CPA_PUB}%
  \BibitemOpen
  \bibfield  {author} {\bibinfo {author} {\bibfnamefont {D.}~\bibnamefont {Strickland}}\ and\ \bibinfo {author} {\bibfnamefont {G.}~\bibnamefont {Mourou}},\ }\bibfield  {title} {\bibinfo {title} {{C}ompression of amplified chirped optical pulses},\ }\href@noop {} {\bibfield  {journal} {\bibinfo  {journal} {Opt. Commun.}\ }\textbf {\bibinfo {volume} {56}},\ \bibinfo {pages} {219} (\bibinfo {year} {1985})}\BibitemShut {NoStop}%
\bibitem [{\citenamefont {Horsley}\ \emph {et~al.}(2023)\citenamefont {Horsley}, \citenamefont {Galiffi},\ and\ \citenamefont {Wang}}]{Horsley2023_Eigenpulses_PUB}%
  \BibitemOpen
  \bibfield  {author} {\bibinfo {author} {\bibfnamefont {S.~A.~R.}\ \bibnamefont {Horsley}}, \bibinfo {author} {\bibfnamefont {E.}~\bibnamefont {Galiffi}},\ and\ \bibinfo {author} {\bibfnamefont {Y.-T.}\ \bibnamefont {Wang}},\ }\bibfield  {title} {\bibinfo {title} {{E}igenpulses of dispersive time-varying media},\ }\href@noop {} {\bibfield  {journal} {\bibinfo  {journal} {Phys. Rev. Lett.}\ }\textbf {\bibinfo {volume} {130}},\ \bibinfo {pages} {203803} (\bibinfo {year} {2023})}\BibitemShut {NoStop}%
\bibitem [{\citenamefont {Ostrovski{\v{i}}}\ and\ \citenamefont {Solomin}(1967)}]{Ostrovskii1967_Inter_PUB}%
  \BibitemOpen
  \bibfield  {author} {\bibinfo {author} {\bibfnamefont {L.~A.}\ \bibnamefont {Ostrovski{\v{i}}}}\ and\ \bibinfo {author} {\bibfnamefont {B.}~\bibnamefont {Solomin}},\ }\bibfield  {title} {\bibinfo {title} {{C}orrect formulation of the problem of wave interaction with a moving parameter jump},\ }\href@noop {} {\bibfield  {journal} {\bibinfo  {journal} {Radiophys. Quantum Electron}\ }\textbf {\bibinfo {volume} {10}},\ \bibinfo {pages} {666} (\bibinfo {year} {1967})}\BibitemShut {NoStop}%
\bibitem [{\citenamefont {Deck-L{\'e}ger}\ and\ \citenamefont {Caloz}(2019)}]{Deck-Leger2019_Inter_CONF}%
  \BibitemOpen
  \bibfield  {author} {\bibinfo {author} {\bibfnamefont {Z.-L.}\ \bibnamefont {Deck-L{\'e}ger}}\ and\ \bibinfo {author} {\bibfnamefont {C.}~\bibnamefont {Caloz}},\ }\bibfield  {title} {\bibinfo {title} {{S}cattering at interluminal interface},\ }in\ \href@noop {} {\emph {\bibinfo {booktitle} {2019 IEEE International Symposium on Antennas and Propagation and USNC-URSI Radio Science Meeting}}}\ (\bibinfo {year} {2019})\ pp.\ \bibinfo {pages} {367--368}\BibitemShut {NoStop}%
\bibitem [{Note1()}]{Note1}%
  \BibitemOpen
  \bibinfo {note} {This scattering time may be interpreted as the emission time of the Huygens' source on the interface that produced the scattered field at $\left (z,ct\right )$.}\BibitemShut {Stop}%
\bibitem [{\citenamefont {Deck-L{\'e}ger}\ \emph {et~al.}(2023)\citenamefont {Deck-L{\'e}ger}, \citenamefont {Bahrami}, \citenamefont {Li},\ and\ \citenamefont {Caloz}}]{Deck-Leger2022_FDTD_PUB}%
  \BibitemOpen
  \bibfield  {author} {\bibinfo {author} {\bibfnamefont {Z.-L.}\ \bibnamefont {Deck-L{\'e}ger}}, \bibinfo {author} {\bibfnamefont {A.}~\bibnamefont {Bahrami}}, \bibinfo {author} {\bibfnamefont {Z.}~\bibnamefont {Li}},\ and\ \bibinfo {author} {\bibfnamefont {C.}~\bibnamefont {Caloz}},\ }\bibfield  {title} {\bibinfo {title} {{G}eneralized {F}{D}{T}{D} scheme for the simulation of electromagnetic scattering in moving structures},\ }\href@noop {} {\bibfield  {journal} {\bibinfo  {journal} {Opt. Express.}\ }\textbf {\bibinfo {volume} {31}},\ \bibinfo {pages} {23214} (\bibinfo {year} {2023})}\BibitemShut {NoStop}%
\bibitem [{\citenamefont {Moussa}\ \emph {et~al.}(2023)\citenamefont {Moussa}, \citenamefont {Xu}, \citenamefont {Yin}, \citenamefont {Galiffi}, \citenamefont {Ra'di},\ and\ \citenamefont {Al{\`u}}}]{Alu2023_Temp_Refl_PUB}%
  \BibitemOpen
  \bibfield  {author} {\bibinfo {author} {\bibfnamefont {H.}~\bibnamefont {Moussa}}, \bibinfo {author} {\bibfnamefont {G.}~\bibnamefont {Xu}}, \bibinfo {author} {\bibfnamefont {S.}~\bibnamefont {Yin}}, \bibinfo {author} {\bibfnamefont {E.}~\bibnamefont {Galiffi}}, \bibinfo {author} {\bibfnamefont {Y.}~\bibnamefont {Ra'di}},\ and\ \bibinfo {author} {\bibfnamefont {A.}~\bibnamefont {Al{\`u}}},\ }\bibfield  {title} {\bibinfo {title} {{O}bservation of temporal reflection and broadband frequency translation at photonic time interfaces},\ }\href@noop {} {\bibfield  {journal} {\bibinfo  {journal} {Nat. Phys.}\ }\textbf {\bibinfo {volume} {19}},\ \bibinfo {pages} {863} (\bibinfo {year} {2023})}\BibitemShut {NoStop}%
\bibitem [{\citenamefont {Galiffi}\ \emph {et~al.}(2023)\citenamefont {Galiffi}, \citenamefont {Xu}, \citenamefont {Yin}, \citenamefont {Moussa}, \citenamefont {Ra'di},\ and\ \citenamefont {Al{\`{u}}}}]{Alu2023_Coherent_Wave_Control_PUB}%
  \BibitemOpen
  \bibfield  {author} {\bibinfo {author} {\bibfnamefont {E.}~\bibnamefont {Galiffi}}, \bibinfo {author} {\bibfnamefont {G.}~\bibnamefont {Xu}}, \bibinfo {author} {\bibfnamefont {S.}~\bibnamefont {Yin}}, \bibinfo {author} {\bibfnamefont {H.}~\bibnamefont {Moussa}}, \bibinfo {author} {\bibfnamefont {Y.}~\bibnamefont {Ra'di}},\ and\ \bibinfo {author} {\bibfnamefont {A.}~\bibnamefont {Al{\`{u}}}},\ }\bibfield  {title} {\bibinfo {title} {{B}roadband coherent wave control through photonic collisions at time interfaces},\ }\href@noop {} {\bibfield  {journal} {\bibinfo  {journal} {Nat. Phys.}\ }\textbf {\bibinfo {volume} {19}},\ \bibinfo {pages} {1703} (\bibinfo {year} {2023})}\BibitemShut {NoStop}%
\bibitem [{\citenamefont {Jones}\ \emph {et~al.}(2024)\citenamefont {Jones}, \citenamefont {Kildishev}, \citenamefont {Segev},\ and\ \citenamefont {Peroulis}}]{Peroulis2024_Time_Ref_TEM_PUB}%
  \BibitemOpen
  \bibfield  {author} {\bibinfo {author} {\bibfnamefont {T.~R.}\ \bibnamefont {Jones}}, \bibinfo {author} {\bibfnamefont {A.~V.}\ \bibnamefont {Kildishev}}, \bibinfo {author} {\bibfnamefont {M.}~\bibnamefont {Segev}},\ and\ \bibinfo {author} {\bibfnamefont {D.}~\bibnamefont {Peroulis}},\ }\bibfield  {title} {\bibinfo {title} {{T}ime-reflection of microwaves by a fast optically-controlled time-boundary},\ }\href@noop {} {\bibfield  {journal} {\bibinfo  {journal} {Nat. Commun.}\ }\textbf {\bibinfo {volume} {15}},\ \bibinfo {pages} {6786} (\bibinfo {year} {2024})}\BibitemShut {NoStop}%
\bibitem [{\citenamefont {Zhou}\ \emph {et~al.}(2020)\citenamefont {Zhou}, \citenamefont {Alam}, \citenamefont {Karimi}, \citenamefont {Upham}, \citenamefont {Reshef}, \citenamefont {Liu}, \citenamefont {Willner},\ and\ \citenamefont {Boyd}}]{Boyd2020_Time_Refr_ENZ_TEM_PUB}%
  \BibitemOpen
  \bibfield  {author} {\bibinfo {author} {\bibfnamefont {Y.}~\bibnamefont {Zhou}}, \bibinfo {author} {\bibfnamefont {M.~Z.}\ \bibnamefont {Alam}}, \bibinfo {author} {\bibfnamefont {M.}~\bibnamefont {Karimi}}, \bibinfo {author} {\bibfnamefont {J.}~\bibnamefont {Upham}}, \bibinfo {author} {\bibfnamefont {O.}~\bibnamefont {Reshef}}, \bibinfo {author} {\bibfnamefont {C.}~\bibnamefont {Liu}}, \bibinfo {author} {\bibfnamefont {A.~E.}\ \bibnamefont {Willner}},\ and\ \bibinfo {author} {\bibfnamefont {R.~W.}\ \bibnamefont {Boyd}},\ }\bibfield  {title} {\bibinfo {title} {{B}roadband frequency translation through time refraction in an epsilon-near-zero material},\ }\href@noop {} {\bibfield  {journal} {\bibinfo  {journal} {Nat. Commun.}\ }\textbf {\bibinfo {volume} {11}},\ \bibinfo {pages} {2180} (\bibinfo {year} {2020})}\BibitemShut {NoStop}%
\bibitem [{\citenamefont {Lustig}\ \emph {et~al.}(2023)\citenamefont {Lustig}, \citenamefont {Segal}, \citenamefont {Saha}, \citenamefont {Bordo}, \citenamefont {Chowdhury}, \citenamefont {Sharabi}, \citenamefont {Fleischer}, \citenamefont {Boltasseva}, \citenamefont {Cohen}, \citenamefont {Shalaev},\ and\ \citenamefont {Segev}}]{Segev2023_Single_Cycle_TEM_PUB}%
  \BibitemOpen
  \bibfield  {author} {\bibinfo {author} {\bibfnamefont {E.}~\bibnamefont {Lustig}}, \bibinfo {author} {\bibfnamefont {O.}~\bibnamefont {Segal}}, \bibinfo {author} {\bibfnamefont {S.}~\bibnamefont {Saha}}, \bibinfo {author} {\bibfnamefont {E.}~\bibnamefont {Bordo}}, \bibinfo {author} {\bibfnamefont {S.~N.}\ \bibnamefont {Chowdhury}}, \bibinfo {author} {\bibfnamefont {Y.}~\bibnamefont {Sharabi}}, \bibinfo {author} {\bibfnamefont {A.}~\bibnamefont {Fleischer}}, \bibinfo {author} {\bibfnamefont {A.}~\bibnamefont {Boltasseva}}, \bibinfo {author} {\bibfnamefont {O.}~\bibnamefont {Cohen}}, \bibinfo {author} {\bibfnamefont {V.~M.}\ \bibnamefont {Shalaev}},\ and\ \bibinfo {author} {\bibfnamefont {M.}~\bibnamefont {Segev}},\ }\bibfield  {title} {\bibinfo {title} {{T}ime-refraction optics with single cycle modulation},\ }\href@noop {} {\bibfield  {journal} {\bibinfo  {journal} {Nanophotonics}\ }\textbf {\bibinfo {volume} {12}},\ \bibinfo {pages} {2221} (\bibinfo {year} {2023})}\BibitemShut {NoStop}%
\bibitem [{\citenamefont {Ball}\ \emph {et~al.}(2025)\citenamefont {Ball}, \citenamefont {Secondo}, \citenamefont {Fomra}, \citenamefont {Wu}, \citenamefont {Saha}, \citenamefont {Agrawal}, \citenamefont {Lezec},\ and\ \citenamefont {Kinsey}}]{Kinsey2025_ST_Knife_TEM_PUB}%
  \BibitemOpen
  \bibfield  {author} {\bibinfo {author} {\bibfnamefont {A.}~\bibnamefont {Ball}}, \bibinfo {author} {\bibfnamefont {R.}~\bibnamefont {Secondo}}, \bibinfo {author} {\bibfnamefont {D.}~\bibnamefont {Fomra}}, \bibinfo {author} {\bibfnamefont {J.}~\bibnamefont {Wu}}, \bibinfo {author} {\bibfnamefont {S.}~\bibnamefont {Saha}}, \bibinfo {author} {\bibfnamefont {A.}~\bibnamefont {Agrawal}}, \bibinfo {author} {\bibfnamefont {H.}~\bibnamefont {Lezec}},\ and\ \bibinfo {author} {\bibfnamefont {N.}~\bibnamefont {Kinsey}},\ }\bibfield  {title} {\bibinfo {title} {{A} space-time knife-edge in epsilon-near-zero films for ultrafast pulse characterization},\ }\href@noop {} {\bibfield  {journal} {\bibinfo  {journal} {Laser Photonics Rev.}\ }\textbf {\bibinfo {volume} {19}},\ \bibinfo {pages} {2401462} (\bibinfo {year} {2025})}\BibitemShut {NoStop}%
\end{thebibliography}%

\end{document}